\documentclass[12pt]{article}
\usepackage{a4wide}
\usepackage{amsmath}
\usepackage{amssymb}
\usepackage{dsfont}
\usepackage{subcaption}
\usepackage{graphicx}
\usepackage{listings}
\usepackage{easybmat}
\usepackage{url}
\usepackage{hyperref}
\usepackage[numbers,sort&compress]{natbib}

\numberwithin{equation}{section}
\numberwithin{figure}{section}

\newcommand{\M}{\mathds{M}}
\newcommand{\A}{\mathds{A}}
\newcommand{\B}{\mathds{B}}
\newcommand{\C}{\mathds{C}}
\newcommand{\D}{\mathds{D}}
\newcommand{\E}{\mathds{E}}
\newcommand{\F}{\mathds{F}}
\newcommand{\G}{\mathds{G}}
\newcommand{\T}{\mathds{T}}
\newcommand{\id}{\mathds{1}}
\renewcommand{\P}{\mathds{P}}
\newcommand{\Pbar}{\overline\P}
\newcommand{\Q}{\mathds{Q}}
\newcommand{\Qbar}{\overline\Q}

\DeclareMathOperator{\rank}{rank}

\newcommand{\uoption}[2]{{\makebox[18em][l]{\texttt{#1}:}\parbox[t]{\dimexpr\linewidth-18em\relax}{#2}}\vspace{1ex}}

\title{\vskip-3cm{\baselineskip14pt
  \begin{flushleft}
      \normalsize TTK-17-01
  \end{flushleft}}
  \vskip1.5cm
	{\tt epsilon}: A tool to find a canonical basis of\\ master integrals
}

\author{\small
  Mario Prausa
  \\[1em]
  {\small \it Institute for Theoretical Particle Physics and Cosmology}\\
  {\small \it RWTH Aachen University}\\
  {\small \it 52056 Aachen, Germany}\\[.5em]
  {\small \tt prausa@physik.rwth-aachen.de}
}  

\date{}

\begin{document}
  \maketitle
  \begin{abstract}
    In 2013, Henn proposed a special basis for a certain class of master integrals, which are expressible in terms of iterated integrals.
    In this basis, the master integrals obey a differential equation, where the right hand side is proportional to $\epsilon$ in $d=4-2\epsilon$ space-time dimensions.
    An algorithmic approach to find such a basis was found by Lee.
    We present the tool \texttt{epsilon}, an efficient implementation of Lee's algorithm based on the \texttt{Fermat} computer algebra system as computational backend.
  \end{abstract}

  {\em Keywords:} Feynman integral; canonical basis; differential equation; Fuchsian form 

  \subsection*{Program Summary}
		\begin{small}
			{\em Program Title:} epsilon \\
			{\em Licensing provisions:} GPLv3 \\
			{\em Programming language:} C++ \\
			{\em Nature of problem:} 
        For a certain class of master integrals, a canonical basis can be found in which they fulfill a differential equation with the right hand side proportional to $\epsilon$.
        In such a basis the solution of the master integrals in an $\epsilon$-expansion becomes trivial.
				Unfortunately, the problem of finding a canonical basis is challenging.
				\\
			{\em Solution method:} Algorithm by Lee~[1] \\
			{\em Restrictions:} 
				The normalization step of Lee's algorithm will fail if the eigenvalues of the matrix residues are not of the form $a+b\epsilon$ with $a,b \in\mathds{Z}$.
				Multi-scale problems are not supported.
        \\[8pt]
        \;[1] R. N. Lee, JHEP 1504 (2015) 108 [arXiv:1411.0911 [hep-ph]].
		\end{small}

  \section{Introduction}
    The perturbative treatment of quantum field theories leads quite naturally to the problem of evaluating a large number of multi-loop Feynman diagrams.
    After a tensor reduction the Feynman diagrams can be expressed in an even larger number of scalar Feynman integrals of the form
    \begin{equation} \label{eq:int}
      \int d^dl_1 \dots \int d^dl_L \frac1{D_1^{n_1} \dots D_N^{n_N}}\,,
    \end{equation}
    where $L$ is the number of loops and $d=4-2\epsilon$ the number of space-time dimensions in the context of dimensional regularization.
    The denominators $D_i$ in \eqref{eq:int} are usually of the form $p^2 - m^2$, where $p$ is a linear combination of loop momenta and external momenta and $m$ some mass.

    A standard technique nowadays is the usage of integration-by-parts identities \cite{Tkachov:1981wb,Chetyrkin:1981qh} for the reduction of this large number of Feynman integrals to a rather small set of so-called master integrals.
    These identities provide linear dependences between various Feynman integrals, where the coefficients are rational functions in both the space-time dimension $d$ and the kinematic variables of the problem.

    Many methods were developed to solve these master integrals.
    For an overview see e.g. \cite{Smirnov:2012gma}.
    Among the most successful ones is the method of differential equations which is also based on integration-by-parts reductions\cite{Kotikov:1990kg,Kotikov:1991hm,Kotikov:1991pm}.
    Recently, significant progress was made in this method, when Henn conjectured the existence of a canonical basis for master integrals expressible in terms of iterated integrals \cite{Henn:2013pwa}.
    In this basis the right hand side of the system of differential equations is proportional to $\epsilon = (4-d)/2$.
    If the boundary conditions are known, the solution of the system of differential equations in an $\epsilon$-series becomes trivial.

    Two years ago, Lee proposed an algorithm to automate finding a canonical basis~\cite{Lee:2014ioa}.
    A first implementation for this algorithm was presented in \cite{Gituliar:2016vfa,Gituliar:2017vzm}.

    In this paper we present \texttt{epsilon}, a further implementation of Lee's algorithm based on the \texttt{Fermat}\cite{Lewis:fermat} computer algebra system.
    Our implementation utilizes the explicit dependence of the transformations used by Lee's algorithm on the kinematic variable to reduce the number of variables in intermediate steps.
    Another advantage of our implementation is the support of systems with singularities at complex points using \texttt{Fermat}'s polymod capability.

    In Section~\ref{sect:impl} we introduce some definitions and explain implementation details.
    In Section~\ref{sect:usage} the installation procedure and the usage of \texttt{epsilon} is described.
    In Section~\ref{sect:example} we give a non-trivial example of the usage based on a real three-loop computation.
  \section{Implementation details} \label{sect:impl}
    \subsection{Definitions}
      We consider a set of $N$ master integrals $\vec{f}$ fulfilling an ordinary system of differential equations
      \begin{equation} \label{eq:system}
        \frac{\partial \vec{f}(x,\epsilon)}{\partial x} 
        = 
        \M(x,\epsilon) \vec{f}(x,\epsilon)\,,
      \end{equation}
      where $x$ is a kinematic variable, $\M(x,\epsilon)$ is an $N\times N$-matrix and $\epsilon$ is a regulator in $d = 4-2\epsilon$ dimensions in the context of dimensional regularization.
      We restrict ourselves to the case
      \begin{equation} \label{eq:M}
        \M(x,\epsilon) 
        = 
        \sum\limits_{x_j \in S} 
        \sum\limits_{k\geq0} 
        \frac{\M^{(x_j)}_k(\epsilon)}{(x-x_j)^{k+1}} 
        + 
        \sum\limits_{k\geq 0} 
        x^k 
        \M_k(\epsilon)\,,
      \end{equation}
      where $S$ is the set of all finite singularities and $\M^{(x_j)}_k$ and $\M_k(\epsilon)$ are independent of $x$.
      In particular, singularities $x_j$ depending on $\epsilon$ are forbidden.
      In many physically relevant cases one can use a trial and error approach to find a basis of master integrals $\vec{f}$ fulfilling the restriction \eqref{eq:M}.
      The main strategy of our implementation is to keep the system always in the form of \eqref{eq:M} since here the $x$-dependence is explicit.

      A singularity $x_j<\infty$ has Poincar\'{e} rank $p$ if $\M^{(x_j)}_p \neq 0$ and $\M^{(x_j)}_k = 0$ for $k>p$.
      In addition to the finite singularities, the system might also have a singularity at $\infty$.
      The Poincar\'{e} rank $p$ of a singularity at $\infty$ is defined as the Poincar\'{e} rank of the singularity at $y=0$ of the system $\M(1/y,\epsilon)/y^2$.
      So \eqref{eq:M} has Poincar\'{e} rank $p>0$ at $\infty$ if $\M_{p-1} \neq 0$ and $\M_k = 0$ for $k\geq p$, and Poincar\'{e} rank $p=0$ at $\infty$ if all $\M_k = 0$ and $\sum_{x_j\in S} \M^{(x_j)}_0 \neq 0$.
      If all $\M_k = 0$ and $\sum_{x_j\in S} \M^{(x_j)}_0 = 0$, the system is not singular at $\infty$.

      Let $p$ be the Poincar\'{e} rank of a singularity $x_j<\infty$, then the generalized Poincar\'{e} rank (or Moser rank) \cite{Moser:1959} of this singularity is defined as $p + r/n - 1$, where $r = \rank \M^{(x_j)}_p$ and $n$ is the dimension of the system.

      A system
      \begin{equation} \label{eq:fuchsian}
        \M(x,\epsilon) 
        = 
        \sum\limits_{x_j \in S} 
        \frac{\M^{(x_j)}_0(\epsilon)}{x-x_j}\,,
      \end{equation}
      where all singularities have Poincar\'{e} rank zero is called Fuchsian, and a system
      \begin{equation} \label{eq:epform}
        \M(x,\epsilon) 
        = 
        \epsilon \sum\limits_{x_j \in S} 
        \frac{\widehat\M^{(x_j)}_0}{x-x_j}\,,
      \end{equation}
      where $\widehat\M^{(x_j)}_0$ is no longer a function of $\epsilon$, is said to be in $\epsilon$-form.
      A change of basis
      \[
        \vec{g}(x,\epsilon) = \T^{-1}(x,\epsilon) \vec{f}
      \]
      modifies the system \eqref{eq:system} to 
      \[
        \frac{\partial \vec{g}(x,\epsilon)}{\partial x} 
        = 
        \widetilde\M(x,\epsilon) \vec{g}(x,\epsilon)\,,
      \]
      with
      \begin{equation} \label{eq:trans}
        \widetilde\M(x,\epsilon)
        =
        \T^{-1}(x,\epsilon)
        \M(x,\epsilon)
        \T(x,\epsilon)
        -
        \T^{-1}(x,\epsilon)
        \frac{\partial}{\partial x}
        \T(x,\epsilon)\,.
      \end{equation}
      We assume the master integrals in $\vec{f}$ to be ordered in a way that a block-triangular structure of the system is obtained (for details see e.g.~\cite{Lee:2014ioa}).
      We will often make use of this block-triangular structure.
      Therefore we write
      \begin{equation} \label{eq:triangular}
        \M
        =
        \begin{pmatrix}
          \A & 0 & 0 \\
          \B & \C & 0 \\
          \D & \E & \F
        \end{pmatrix}\,,
      \end{equation}
      and use the same indices as in \eqref{eq:M} for the matrices $\A,\dots,\F$ (e.g. $\C^{(x_j)}_k(\epsilon)$).
      The block $\C$ is called the active block as we apply Lee's algorithm to this block.
      As $\A$ to $\F$ are matrices as well, the definition of what we call the active block is more or less arbitrary as long as a block-triangular structure is obtained.
      But from a computational point of view a small dimension of the active block is preferable since this reduces the complexity of the resulting operations.
      In the following, the matrices $\A^{(x_j)}_k, \dots, \F^{(x_j)}_k$ and $\A_k,\dots,\F_k$ will be referred to as coefficient matrices.
    \subsection{Utilizing the explicit $x$-dependence}
      Lee's algorithm uses three types of transformations: balances, off-diagonal reductions and $x$-independent transformations.

      We define balances as
      \begin{subequations} \label{eq:bal}
        \begin{align}
          {\cal B}(\P,x_1,x_2) &= \Pbar + \frac{x-x_2}{x-x_1} \P\,, \label{eq:bal_x1_x2} \\
          {\cal B}(\P,x_1,\infty) &= \Pbar + \frac1{x-x_1} \P\,, \label{eq:bal_x1_inf} \\
          {\cal B}(\P,\infty,x_2) &= \Pbar + (x-x_2) \P\,, \label{eq:bal_inf_x2}
        \end{align}
      \end{subequations}        
      where $\P$ is a projector to be specified below, depending only on $\epsilon$, $\Pbar = \id - \P$ and $x_1,x_2 < \infty$.
      In Lee's algorithm balances are applied to the active block $\C$ in order to reduce the generalized Poincar\'{e} rank of singular points and to normalize eigenvalues of Fuchsian singularities.
      
      Off-diagonal reductions are used to reduce the block $\B$ to Fuchsian form after the blocks $\A$ and $\C$ were already reduced to $\epsilon$-form.
      They are defined by
      \begin{subequations} \label{eq:offred}
        \begin{align}
          {\cal L}(x_1,k,\G) &= \id + \frac1{(x-x_1)^k} \G\,, \label{eq:offred1} \\
          {\cal L}(\infty,k,\G) &= \id + x^k \G\,, \label{eq:offred2}
        \end{align}
        where
        \begin{equation} \label{eq:offredG}
          \G =
          \begin{pmatrix}
            0 & 0 & 0 \\ 
            \widehat\G & 0 & 0 \\ 
            0 & 0 & 0 
          \end{pmatrix}\,.
        \end{equation}
      \end{subequations}        
      The block $\widehat\G$ has the same boundaries in $\G$ as block $\B$ in \eqref{eq:triangular}. Note that $\G^2 = 0$.

      In both types of transformations the $x$-dependence is explicit.
			Another type of transformation which is independent of $x$ is used in the last step of Lee's algorithm to factor out $\epsilon$.
			
      Our goal is to use those three types in the transformation rule \eqref{eq:trans} without spoiling the form \eqref{eq:M} or the block-triangular structure \eqref{eq:triangular} of the system.
     
      As a pedagogical example we consider the transformation of block $\B$ under a balance between two singularities $x_1$ and $x_2$, i.e.
      \begin{equation} \label{eq:bal_ex_trans}
        \T 
        = 
        {\cal B}(\P,x_1,x_2)
        =
        \Pbar + \frac{x-x_2}{x-x_1} \P\,, \quad
        \T^{-1}
        =
        {\cal B}(\P,x_2,x_1)
        =
        \Pbar + \frac{x-x_1}{x-x_2} \P\,.
      \end{equation}
			Since we want to apply Lee's algorithm to the active block we can restrict the form of the projector $\P$ to 
      \begin{equation} \label{eq:Qdef}
        \P = \begin{pmatrix}
          0 & 0 & 0 \\
          0 & \Q & 0 \\
          0 & 0 & 0
        \end{pmatrix}\,,
      \end{equation}
      where $\Q$ is a projector with the dimensions of the active block.
      Inserting \eqref{eq:bal_ex_trans} into \eqref{eq:trans} we obtain
      \begin{align*}
        \widetilde\M(x,\epsilon)
        &=
        \left[
          \Pbar + \frac{x-x_1}{x-x_2} \P
        \right]          
        \M(x,\epsilon)
        \left[
          \Pbar + \frac{x-x_2}{x-x_1} \P
        \right] 
        +
        \frac{x_2-x_1}{(x-x_1)(x-x_2)}
        \P \\
        &=
        \M(x,\epsilon)
        -
        \P \M(x,\epsilon) \Pbar
        -
        \Pbar \M(x,\epsilon) \P
        +
        \frac{x-x_2}{x-x_1} \Pbar \M(x,\epsilon) \P
        +
        \frac{x-x_1}{x-x_2} \P \M(x,\epsilon) \Pbar \\ &\quad
        +
        \frac{x_2-x_1}{(x-x_1)(x-x_2)}
        \P 
      \end{align*}
      So block $\B$ in \eqref{eq:triangular} transforms as
      \[
        \widetilde\B(x,\epsilon)
        =
        \B(x,\epsilon)
        -
        \Q \B(x,\epsilon)
        +
        \frac{x-x_1}{x-x_2} \Q \B(x,\epsilon)\,.
      \]
     Inserting the form \eqref{eq:M} of $\B(x,\epsilon)$ leads to
      \begin{equation} \label{eq:tildeB}
        \widetilde\B(x,\epsilon)
        =
        \B(x,\epsilon)
        -
        \Q \B(x,\epsilon)
        +
        \sum\limits_{x_j \in S} 
        \sum\limits_{k\geq0} 
        \frac{(x-x_1) \Q \B^{(x_j)}_k(\epsilon)}{(x-x_2)(x-x_j)^{k+1}} 
        + 
        \sum\limits_{k\geq 0} 
        \frac{(x-x_1)x^k}{x-x_2} 
        \Q \B_k(\epsilon)\,.
      \end{equation}
      Using partial fractioning and the incomplete geometric series, we can show that
			\begin{subequations} \label{eq:auxidentities}
				\begin{align}
					\begin{split} \label{eq:auxid1}
						\sum\limits_{k=0}^\infty 
						\frac{a_k}{(x-x_2)(x-x_j)^{k+1}}
						&=
						\frac1{x-x_2} 
						\sum\limits_{n\geq0} 
						\frac{a_n}{(x_2-x_j)^{n+1}} \\ &\quad 
						- 
						\sum\limits_{k\geq0} 
						\frac1{(x-x_j)^{k+1}} 
						\sum\limits_{n=0}^\infty 
						\frac{a_{n+k}}{(x_2-x_j)^{n+1}} \,,
					\end{split} \\
					\sum\limits_{k\geq0} 
					\frac{x^k}{x-x_2} a_k 
					&=
					\sum\limits_{k\geq0} 
					x^k 
					\sum\limits_{n\geq0} 
					x_2^n \;
					a_{k+n+1} 
					+ 
					\frac1{x-x_2} 
					\sum\limits_{n\geq0} 
					x_2^n \; a_n \,,
				\end{align}
			\end{subequations}
      where identity \eqref{eq:auxid1} only holds for $x_j\neq x_2$.
      Combining \eqref{eq:tildeB} and \eqref{eq:auxidentities} yields
      \begin{equation} \label{eq:tildeB2}
				\begin{split}
					\widetilde\B(x,\epsilon)
					&=
					\B(x,\epsilon)
					+
					\frac{x_2-x_1}{x-x_2} 
					\sum\limits_{x_j \in S\backslash\{x_2\}} 
					\sum\limits_{n\geq0} 
					\frac{\Q \B^{(x_j)}_n}{(x_2-x_j)^{n+1}}
					+ 
					\frac{x_2-x_1}{x-x_2} 
					\sum\limits_{n\geq0} 
					x_2^n \Q \B_n \\ &\quad 
					+
					\sum\limits_{k\geq1} 
					\frac{(x_2-x_1) \Q \B^{(x_2)}_{k-1}(\epsilon)}{(x-x_2)^{k+1}}
					+ 
					\sum\limits_{x_j \in S\backslash\{x_2\}} 
					\sum\limits_{k\geq0} 
					\frac{x_1-x_2}{(x-x_j)^{k+1}} 
					\sum\limits_{n\geq0} 
					\frac{\Q \B^{(x_j)}_{n+k}(\epsilon)}{(x_2-x_j)^{n+1}} \\ &\quad
					+
					(x_2-x_1)
					\sum\limits_{k\geq0} 
					x^k 
					\sum\limits_{n\geq0} 
					x_2^n 
					\Q \B_{k+n+1}(\epsilon)\,.
				\end{split}
      \end{equation}
      Hence, the transformation laws for the coefficient matrices can be found by comparing \eqref{eq:tildeB2} with the structure of \eqref{eq:M}:
      \begin{align*}
        \widetilde\B^{(x_2)}_0(\epsilon)
        &=
        \B^{(x_2)}_0(\epsilon)
        +
        \sum\limits_{x_j \in S\backslash\{x_2\}} 
        \sum\limits_{n\geq0} 
        \frac{x_2-x_1}{(x_2-x_j)^{n+1}}
        \Q \B^{(x_j)}_n(\epsilon)
        +
        (x_2-x_1)
        \sum\limits_{n\geq0} 
        x_2^n \Q \B_n(\epsilon)
        \,, \\
        \widetilde\B^{(x_2)}_{k>0}(\epsilon)
        &=
        \B^{(x_2)}_k(\epsilon)
        +
        (x_2-x_1) \Q \B^{(x_2)}_{k-1}(\epsilon)
        \,, \\
        \widetilde\B^{(x_j\neq x_2)}_k(\epsilon)
        &=
        \B^{(x_j)}_k(\epsilon)
        +
        \sum\limits_{n\geq0} 
        \frac{x_1-x_2}{(x_2-x_j)^{n+1}}
        \Q \B^{(x_j)}_{n+k}(\epsilon)
        \,, \\
        \widetilde\B_k(\epsilon)
        &=
        (x_2-x_1)
        \sum\limits_{n\geq0} 
        x_2^n 
        \Q \B_{k+n+1}(\epsilon)
        \,.
      \end{align*}
      An advantage of this form over the original transformation \eqref{eq:bal_ex_trans} is that now all operations are independent of $x$.
      Therefore, the underlying computer algebra system has to deal with rational functions of one less variable.
      The form \eqref{eq:M} remains unspoiled, i.e. it is not necessary to perform a partial fraction decomposition after the transformation.

      All transformations in terms of the coefficient matrices are listed in appendix \ref{sect:app_bal}.
    \subsection{Overview of Lee's algorithm} \label{sect:lee}
      Three basic steps allow Lee's algorithm\cite{Lee:2014ioa} to transform an ordinary system of differential equations into an $\epsilon$-form \eqref{eq:epform} if they are applied to the whole system:
      \begin{enumerate}
        \item transformation of a system into Fuchsian form,
        \item normalization of the eigenvectors of all matrix residues,
        \item factorization of $\epsilon$.
      \end{enumerate}
      In order to make use of the block-triangular structure of the system \eqref{eq:triangular} these three steps are applied only to the active block followed by a fourth step to transform the off-diagonal block $\B$ into Fuchsian form.

      In this sub-section we briefly discuss all four steps.
      More details can be found in the original paper by Lee\cite{Lee:2014ioa}.

      \subsubsection*{Fuchsification}
        The basic building blocks for the first part of Lee's algorithm, the transformation of the system to Fuchsian form, are the balances defined in \eqref{eq:bal}.
        With the right choice of a projector $\P$, it is possible to perform a so-called Moser reduction to strictly lower the generalized Poincar\'{e} rank of the singularity at $x_1$~\cite{Moser:1959}.
        In this discussion we restrict ourselves to the case where $x_1<\infty$.
        A more general treatment is given in\cite{Lee:2014ioa}.
        
        Let $p$ be the Poincar\'{e} rank of the singularity at $x_1$ of the active block $\C$ and $\{u_k^{\alpha}\}$ with $\alpha = 0,\dots,n_k$ be the set of $n_k+1$ right generalized eigenvectors of $\C_p^{(x_1)}$ belonging to a Jordan block $k$ in the Jordan decomposition of $\C_p^{(x_1)}$.
        The Jordan blocks are ordered by their size so that $n_i \geq n_{i+1}$.
        If $p>0$, we assume all eigenvalues of $\C_p^{(x_1)}$ to be zero, or else no transformation to lower the generalized Poincar\'{e} rank exists.
        The right generalized eigenvectors fulfill
        \[
          \C_p^{(x_1)} u_k^{(0)} = 0\,, \quad
          \C_p^{(x_1)} u_k^{(\alpha+1)} = u_k^{(\alpha)}\,.
        \]
        These relations are invariant under the transformation
        \begin{equation} \label{eq:eigen_trans}
          u_k^{(\alpha)} \rightarrow u_k^{(\alpha)} + cu_l^{(\alpha)}\,,
        \end{equation}
        where $\alpha=0,\dots,n_k$ and $k>l$.
        The left generalized eigenvectors $v_k^{(\alpha)}$ are related to the right generalized eigenvectors by
        \begin{equation} \label{eq:left_eigen}
          \left(
            v_1^{(n_1)},\dots,v_1^{(0)},
            v_2^{(n_2)},\dots,v_2^{(0)},
            \dots
          \right)
          =
          \left[
            \left(
              u_1^{(0)},\dots,u_1^{(n_1)},
              u_2^{(0)},\dots,u_2^{(n_2)},
              \dots
            \right)^{-1}
          \right]^\dagger\,,
        \end{equation}
        and fulfill
        \[
          v_k^{(0)\dagger} \C_p^{(x_1)} = 0\,, \quad
          v_k^{(\alpha+1)\dagger} \C_p^{(x_1)} = v_k^{(\alpha)\dagger}\,.
        \]
        The transformation \eqref{eq:eigen_trans} allows us to find a basis of eigenvectors which satisfies
        \begin{equation} \label{eq:eigen_cond}
          v_j^{(0)\dagger} \C_{p-1}^{(x_1)} u_k^{(0)} = 0\,,
        \end{equation}
        for $j \notin R$ and $k \in R \cup \{k_0\}$, where $R$ is some set of trivial Jordan blocks ($n_i = 0$ for $i\in R$) and $k_0$ is a non-trivial Jordan block ($n_{k_0} > 0$).
        An algorithm to find these generalized eigenvectors, together with the set $R$ and $k_0$ is given in~\cite{Lee:2014ioa}.

        If in the definition of $\P$ \eqref{eq:Qdef} we use
        \begin{equation} \label{eq:Q_fuchs}
          \Q = \sum\limits_{k\in R\cup\{k_0\}} u_k^{(0)} w_k^\dagger\,,
        \end{equation}
        with $w_j^\dagger u_k^{(0)} = \delta_{jk}$ and then apply the balance \eqref{eq:bal_x1_x2}, the generalized Poincar\'{e} rank of the singularity at $x_1$ is decreased.
        To see this, we introduce the projector
        \begin{equation} \label{eq:Q1_fuchs}
          \Q_1 = \sum\limits_{k\in R\cup\{k_0\}} u_k^{(0)} v_k^{(n_k)\dagger}\,.
        \end{equation}
        Since $v_j^{(n_j)\dagger} u_k^{(0)} = \delta_{jk}$, the projector $\Q_1$ also is of the form \eqref{eq:Q_fuchs}.
        From the definitions of $\Q$ and $\Q_1$ follows
        \[
          \Q_1 \Q = \Q\,, \quad
          \Q \Q_1 = \Q_1\,, \quad
          \C^{(x_1)}_p \Q = \C^{(x_1)}_p \Q_1 = 0\,.
        \]          
        Evaluating the transformation \eqref{eq:bal_x1_x2_Cx1k} at $k=p$ leads to 
        \begin{align*}
          \widetilde \C^{(x_1)}_p &=
          \Qbar \C^{(x_1)}_p
          +
          (x_1-x_2) \Qbar \C^{(x_1)}_{p-1} \Q \\
          &=
          \left(\Qbar+\Q_1\right) 
          \left[
            \Qbar_1 \C^{(x_1)}_p
            +
            (x_1-x_2) \Qbar_1 \C^{(x_1)}_{p-1} \Q_1
          \right]
          \left(\Qbar_1+\Q\right)
        \end{align*} 
        As $(\Qbar+\Q_1) = (\Qbar_1+\Q)^{-1}$, the matrix rank of $\widetilde \C^{(x_1)}_p$ is given by
        \[
          \rank \widetilde \C^{(x_1)}_p = 
          \rank \widehat\C^{(x_1)}_p\,, \quad
          \widehat\C^{(x_1)}_p =
          \Qbar_1 \C^{(x_1)}_p
          +
          (x_1-x_2) \Qbar_1 \C^{(x_1)}_{p-1} \Q_1,.
        \]
				The argument that the matrix rank (and therefore the generalized Poincar\'{e} rank) of $\widetilde \C^{(x_1)}_p$ is lower than the matrix rank of $\C^{(x_1)}_p$ is as follows:
        \begin{itemize}
          \item
            All left eigenvectors $v_j^{(0)}$ of $\C^{(x_1)}_p$ with $j\in R$ are left eigenvectors of $\widehat \C^{(x_1)}_p$ as $v_j^{(0)\dagger} \Qbar_1 = 0$.
          \item
            All left eigenvectors $v_j^{(0)}$ of $\C^{(x_1)}_p$ with $j\notin R$ are left eigenvectors of $\widehat \C^{(x_1)}_p$ as $v_j^{(0)\dagger} \Qbar_1 = v_j^{(0)\dagger}$ and so
            \[
              v_j^{(0)\dagger} \widehat\C^{(x_1)}_p 
              =
              (x_1-x_2) \sum\limits_{k\in R\cup\{k_0\}} v_j^{(0)\dagger} \C^{(x_1)}_{p-1} u_k^{(0)} v_k^{(n_k)\dagger}
              = 0\,,
            \]
            where we used \eqref{eq:eigen_cond}.
          \item
            The vector $v_{k_0}^{(n_{k_0})}$ which is not a left eigenvector of $\C^{(x_1)}_p$ is an additional left eigenvector of $\widehat \C^{(x_1)}_p$ as $v_{k_0}^{(n_{k_0})\dagger} \Qbar_1 = 0$.
        \end{itemize}
        So $\widehat \C^{(x_1)}_p$ has one eigenvector more than $\C^{(x_1)}_p$ and therefore has a lower matrix rank.
        
        Unfortunately, the balance \eqref{eq:bal_x1_x2} might also increase the Poincar\'{e} rank at $x_2$.
				Therefore, we have a closer look at \eqref{eq:bal_x1_x2_Cx2k} evaluated at $k=q+1$, where $q$ is the Poincar\'{e} rank at the singularity $x_2$:
        \[
          \widetilde \C^{(x_2)}_{q+1} 
          =
          (x_2-x_1)
          \Q \C^{(x_2)}_q \Qbar\,.
        \]
        If this expression vanishes, the Poincar\'{e} rank at $x_2$ is not increased.
        This is the case if the vectors $w_k^\dagger$ in \eqref{eq:Q_fuchs} span a left-invariant space of $\C^{(x_2)}_q$.
        
				In Lee's algorithm the above steps are used to decrease the Poincar\'{e} rank of a singularity $x_1$ as long as it is possible to find a projector \eqref{eq:Q_fuchs} with $w_k^\dagger$ spanning a left-invariant space of some $\C^{(x_2)}_q$, where $x_2 \neq x_1$ is some other singular point.
        If no such projector exists, a regular point $y$ is chosen and the projector $\Q_1$ defined by \eqref{eq:Q1_fuchs} is used in the balance.
        This of course creates a new (apparent) Fuchsian singularity at $y$.
        This way it is possible to reduce the system to an equivalent one with all singularities having Poincar\'{e} rank zero.
        Fortunately, the next step, the normalization of the eigenvalues, removes all apparent singularities introduced in this step.
      \subsubsection*{Normalization}
				The second step of Lee's algorithm is the normalization of the eigenvalues of the matrix residues.
        Again, the main ingredients are the balances defined in \eqref{eq:bal}.
        We assume all singularities to be Fuchsian now and the eigenvalues of the matrix residues to be of the form $a+b\epsilon$ with $a\in\mathds{Z}$.
        This is a necessary condition for the existence of a transformation normalizing all eigenvalues.
				If this condition cannot be fulfilled a redefinition of the kinematic variable might be helpful.
        A normalized eigenvalue is an eigenvalue proportional to $\epsilon$ (i.e. with $a=0$).
        The matrix residue at $\infty$ is equal to $-\sum_{x_j\in S} \C^{(x_j)}_0$.
        Therefore, the sum of all matrix residues (including the residue at $\infty$) vanishes and hence the sum of all eigenvalues of all matrix residues must vanish as well.

        For brevity's sake, we will restrict our discussion to finite singularities $x_1,x_2<\infty$; the generalized treatment can again be found in\cite{Lee:2014ioa}.

        Let $\{u_k^{(\alpha)}\}$ with $\alpha=0,\dots,n_k$ be the set of $n_k+1$ right generalized eigenvectors of $\C^{(x_1)}_0$ belonging to the Jordan block $k$:
        \[
          \C^{(x_1)}_0 u_k^{(0)} = \lambda_k u_k^{(0)}\,, \quad
          \C^{(x_1)}_0 u_k^{(\alpha+1)} = \lambda_k u_k^{(\alpha+1)} + u_k^{(\alpha)}\,.
        \]          
        As in the fuchsification step, we define left generalized eigenvectors $v_k^{(\alpha)}$ via \eqref{eq:left_eigen} which satisfy
        \[
          v_k^{(0)\dagger} \C^{(x_1)}_0 = \lambda_k v_k^{(0)\dagger}\,, \quad
          v_k^{(\alpha+1)\dagger} \C^{(x_1)}_0 = \lambda_k v_k^{(\alpha+1)\dagger} + v_k^{(\alpha)\dagger}\,.
        \]
       
        A balance \eqref{eq:bal_x1_x2} with a right choice of a projector $\P$ can now be used to shift one eigenvalue of the matrix residue at $x_1$ up by one and/or one eigenvalue of the matrix residue at $x_2$ down by one.
        Let us consider a projector $\P$ defined by \eqref{eq:Qdef} with
        \begin{equation} \label{eq:Qnorm}
          \Q = u_k^{(0)} w^\dagger\,,
        \end{equation}
        where $w^\dagger u_k^{(0)} = 1$.
        Also an additional projector
        \[
          \Q_1 = u_k^{(0)} v_k^{(n_k)}
        \]
        is useful for the discussion.
        The following relations involving these two projectors hold:
        \[
          \Q\Q_1 = \Q_1\,, \quad
          \Q_1\Q = \Q\,, \quad
          \C^{(x_1)}_0 \Q = \lambda_k \Q\,, \quad
          \C^{(x_1)}_0 \Q_1 = \lambda_k \Q_1\,.
        \]          
        
        Let us now consider the transformation of $\C^{(x_1)}_0$ under a balance \eqref{eq:bal_x1_x2} as given in \eqref{eq:bal_x1_x2_Cx10}:
        \begin{align*}
          \widetilde \C^{(x_1)}_0 
          &=
          \C^{(x_1)}_0
          -
          \Q \C^{(x_1)}_0 \Qbar 
          +
          \sum\limits_{x_j\in S\backslash\{x_1\}} 
          \frac{x_1-x_2}{x_1-x_j} 
          \Qbar \C^{(x_j)}_0 \Q
          + \Q \\
          &=
          (\Qbar + \Q_1) \widehat \C^{(x_1)}_0 (\Qbar_1 + \Q)\,,
        \end{align*}
        where
        \[
          \widehat \C^{(x_1)}_0 
          =
          \C^{(x_1)}_0
          -
          \Q_1 \C_0^{(x_1)} \Qbar_1
          +
          \sum\limits_{x_j\in S\backslash\{x_1\}} 
          \frac{x_1-x_2}{x_1-x_j} 
          \Qbar_1 \C^{(x_j)}_0 \Q_1
          + \Q_1\,.
        \]            
        Because of $(\Qbar + \Q_1) = (\Qbar_1 + \Q)^{-1}$, $\widetilde \C^{(x_1)}_0$ and $\widehat \C^{(x_1)}_0$ are related by a similarity transformation and thus have the same eigenvalues.
				To evaluate the eigenvalues of $\widetilde \C^{(x_1)}_0$ it is therefore sufficient to analyze the eigenvalues of $\widehat \C^{(x_1)}_0$ which are much simpler to determine.
        We consider $\widehat\C^{(x_1)}_0$ in the basis of the generalized eigenvectors of $\C^{(x_1)}_0$.
        In this basis $\C^{(x_1)}_0$ is in Jordan normal form.
        The second term $-\Q_1 \C_0^{(x_1)} \Qbar_1$ removes all elements from the row corresponding to $u_k^{(0)}$ but the diagonal one containing the eigenvalue to $u_k^{(0)}$.
        The last term $+\Q_1$ increases the diagonal element by one.
        The terms proportional to $\Qbar_1 \C_0^{(x_j)} \Q_1$ contribute to the non-diagonal elements of the column corresponding to $u_k^{(0)}$ (or $v_k^{(n_k)\dagger}$).
        Hence, the transformations can be summarized as 
        \[
          {\footnotesize
            \left(\begin{BMAT}{cccccccccc}{ccccccccc}
              \ddots & & & & & & & & & \\
              & \lambda_{k-1} & & & & & & & & \\
              & & \lambda_k & 1 & & & & & & \\
              & & & \lambda_k & 1  & & & & & \\
              & & & & \ddots & \ddots  & & & &  \\
              & & & & & \lambda_k & 1 & & & \\
              & & & & & & \lambda_k & & & \\
              & & & & & & & \lambda_{k+1} & 1 & \\
              & & & & & & & & \ddots & \ddots
              \addpath{(1,7,.)ru}
              \addpath{(2,7,.)rrrrrdddddllllluuuuu}
              \addpath{(7,1,.)urr}
            \end{BMAT}\right)
            \rightarrow
            \left(\begin{BMAT}{cccccccccc}{ccccccccc}
              \ddots & & \vdots & & & & & & & \\
              & \lambda_{k-1} & \ast & & & & & & & \\
              & & \lambda_k+1 & 0 & & & & & & \\
              & & \ast & \lambda_k & 1  & & & & & \\
              & & \vdots & & \ddots & \ddots  & & & &  \\
              & & \ast & & & \lambda_k & 1 & & & \\
              & & \ast & & & & \lambda_k & & & \\
              & & \ast & & & & & \lambda_{k+1} & 1 & \\
              & & \vdots & & & & & & \ddots & \ddots
              \addpath{(1,7,.)ru}
              \addpath{(2,7,.)rrrrrdddddllllluuuuu}
              \addpath{(7,1,.)urr}
            \end{BMAT}\right)\,,
          }
        \]          
        where $\ast$ stands for contributions from the $\Qbar_1 \C_0^{(x_j)} \Q_1$ terms.
        Calculating the characteristic polynomial by means of a Laplace expansion along the row corresponding to $u_k^{(0)}$ leads to the conclusion that all eigenvalues but one stay the same; only one eigenvalue $\lambda_k$ is changed to $\lambda_k+1$.

        In the same way, a projector
        \begin{equation} \label{eq:Qpnorm}
          \Q' = w v_k^{(0)\dagger}
        \end{equation}
        with a left eigenvector $v_k^{(0)\dagger}$ of $\C^{(x_2)}_0$ and $v_k^{(0)\dagger} w = 1$ shifts one eigenvalue $\lambda_k$ of $\C^{(x_2)}_0$ down by one.
        
        A balance with a projector \eqref{eq:Qnorm} could spoil the Fuchsian form of the system if it increases the Poincar\'{e} rank at any singularity.
        This is in principle possible at $x_2$.
        Evaluating \eqref{eq:bal_x1_x2_Cx2k} at $k=1$ leads to
        \[
          \widetilde \C^{(x_2)}_1 =
          (x_2-x_1)
          \Q \C^{(x_2)}_0 \Qbar\,.
        \]
        This vanishes if $w^\dagger$ is a left eigenvector of $\C^{(x_2)}_0$.
        In that case not only the Fuchsian form of the system is preserved but we also arrive at a projector of the form \eqref{eq:Qpnorm}.
				So the projector of choice is 	
        \begin{equation} \label{eq:Qnorm1}
          \Q = u^{(0)}_k v^{(0)\dagger}_l\,,
        \end{equation}
				where $u^{(0)}_k$ is a right eigenvector of $\C^{(x_1)}_0$, $v^{(0)\dagger}_l$ is a left eigenvector of $\C^{(x_2)}_0$ and $v^{(0)\dagger}_l u^{(0)}_k = 1$.
				This projector, used in balance \eqref{eq:bal_x1_x2}, increases one eigenvalue $\lambda_k$ of $\C^{(x_1)}_0$ by one and decreases one eigenvalue $\mu_l$ of $\C^{(x_2)}_0$ by one while conserving the Fuchsian form of the system.

 			  In order to utilize the considerations above in an algorithmic approach, one first selects a singularity $x_0$ as `fallback'.
        Then balances with projectors of the form \eqref{eq:Qnorm1} are used to `mutually balance' eigenvalues between two singularities. 
        Certainly, the eigenvalue to be increased should be negative and the eigenvalue to be decreased should be positive (for $\epsilon=0$).

        If no such balance exists, the eigenvalues will be balanced with the `fallback singularity' $x_0$ regardless of the sign of the eigenvalue at $x_0$.
        This normalizes the eigenvalues at all singularities but $x_0$.
        To normalize even the eigenvalues at $x_0$, we balance one unnormalized eigenvalue with some regular point creating a new apparent singularity there and restart the algorithm.
        Hopefully, the unnormalized eigenvalue at this new apparent singularity can now be mutually balanced with another unnormalized eigenvalue at $x_0$.

      \subsubsection*{$\epsilon$-Factorization}
				In the next step we find an $x$-independent transformation $\T(\epsilon)$ to factor out $\epsilon$.
        For an $x$-independent $\T$, \eqref{eq:trans} becomes a similarity transformation and does not change the eigenvalues of the system.
				This makes it necessary for the eigenvalues to be proportional to $\epsilon$, i.e. the normalization step before must have been successful.
        We use the fact that $\T^{-1}(\epsilon) [\M^{(x_j)}_0(\epsilon)/\epsilon] \T(\epsilon)$ should be independent of $\epsilon$, so that the equation 
        \[
          \T^{-1}(\epsilon) \frac{\M^{(x_j)}_0(\epsilon)}\epsilon \T(\epsilon)
          =
          \T^{-1}(\mu) \frac{\M^{(x_j)}_0(\mu)}\mu \T(\mu)
        \]
        holds for all $x_j \in S$.
        Multiplying this equation from the left by $\T(\epsilon)$ and from the right by $\T^{-1}(\mu)$, leads to
        \begin{equation} \label{eq:epfact}
          \frac{\M^{(x_j)}_0(\epsilon)}\epsilon \T(\epsilon,\mu)
          =
          \T(\epsilon,\mu) \frac{\M^{(x_j)}_0(\mu)}\mu\,,
        \end{equation}
        where $\T(\epsilon,\mu) = \T(\epsilon)\T^{-1}(\mu)$.
        This linear system can be solved e.g. with Gaussian elimination and the constants should be fixed so that $\T(\epsilon,\mu_0)$ is an invertible matrix, where $\mu_0$ is some arbitrary number.
        The transformation $\T(\epsilon,\mu_0)$ can now be used to factor out $\epsilon$.

      \subsubsection*{Fuchsification of off-diagonal blocks}
				Since the definition of the active block is somewhat arbitrary, we should in principle be able to transform all diagonal blocks to $\epsilon$-form by redefining the active block and applying the three steps described above.
				However, we still need to reduce the off-diagonal block $\B$ to Fuchsian form.
        Again we will restrict the discussion to finite singularities $x_1<\infty$ and assume the block $\A$ and $\C$ to be already in $\epsilon$-form.
       
        Let $p$ be the Poincar\'{e} rank of the off-diagonal block $\B$ at the singularity $x_1$, i.e. $\B^{(x_1)}_p \neq 0$ and $\B^{(x_1)}_k = 0$ for $k>p$.
        The behavior of $\B^{(x_1)}_p$ under a transformation \eqref{eq:offred1} with $k=p$ is given by \eqref{eq:offred_Bx1k}:
        \[
          \widetilde\B^{(x_1)}_p = 
            \B^{(x_1)}_p
            + \C_0^{(x_1)}\widehat\G - \widehat\G\A_0^{(x_1)} 
            + p\widehat\G\,.
        \]
        In order to decrease the Poincar\'{e} rank, $\widehat\G$ has to be determined such that $\widetilde\B^{(x_1)}_p$ vanishes.
        This linear system of equations can be solved e.g. with Gaussian elimination.
       
        Hence, with transformations of the form \eqref{eq:offred} it is possible to reduce all singularities of the off-diagonal block $\B$ to Fuchsian form.
  \newpage        
  \section{Usage} \label{sect:usage}
    \subsection{Installation guide on Linux systems}
      Ensure that the dependencies 
      \begin{itemize}
        \item \texttt{Fermat} ($\geq$ 6.0) \cite{Lewis:fermat},
        \item \texttt{GiNaC} ($\geq$ 1.6.2) \cite{Bauer:2000cp} (for \texttt{epsilon-prepare} only)
      \end{itemize}
      are installed.

      As a next step, \texttt{libFermat} has to be installed.
      \texttt{libFermat}, which was developed in connection with \texttt{epsilon}, is a \texttt{C++} library designed to communicate with \texttt{Fermat}.
      Nevertheless, we decided to publish \texttt{libFermat} in a separate repository since it might be useful elsewhere.
      Internally, the communication is done with \texttt{PStreams}\cite{pstreams} which is included in the package.
      The source code of the most recent version of \texttt{libFermat} can be obtained via \texttt{github} using
      \begin{lstlisting}[basicstyle=\ttfamily,xleftmargin=-4em]
        git clone https://github.com/mprausa/libFermat.git
      \end{lstlisting}
      This will create a directory \texttt{libFermat/} and clone the library into that location.
      Now inside this directory, run
      \begin{lstlisting}[basicstyle=\ttfamily,xleftmargin=-4em]
        cmake -DCMAKE_INSTALL_PREFIX=/path/to/install .
        make
        make install
      \end{lstlisting}
      where \texttt{/path/to/install} is your desired installation directory and defaults to \texttt{/usr/local} on a typical Linux system.
      The library is installed into the sub-directory \texttt{lib} and the header files into the sub-directory \texttt{include} of \texttt{/path/to/install}.
      If your choice is a global directory you will require \texttt{root} access for the last step \texttt{make install}.
      Remember to include the sub-directory \texttt{lib} of \texttt{/path/to/install} into the \texttt{LD\_LIBRARY\_PATH} environment variable if you are using a non-standard directory.
      
      The next step is to install \texttt{epsilon} and \texttt{epsilon-prepare}.
      In principle the procedure is the same as for the installation of \texttt{libFermat}.
      First, obtain the most recent version of the source code with
      \begin{lstlisting}[basicstyle=\ttfamily,xleftmargin=-4em]
        git clone https://github.com/mprausa/epsilon.git
      \end{lstlisting}
      then change into the newly created directory \texttt{epsilon/} and run
      \begin{lstlisting}[basicstyle=\ttfamily,xleftmargin=-4em]
        cmake -DCMAKE_INSTALL_PREFIX=/path/to/install .
        make
        make install
      \end{lstlisting}
      It is recommended to use the same \texttt{/path/to/install} as for \texttt{libFermat}, else the \texttt{cmake} step might require additional options to find \texttt{libFermat}.
      The programs \texttt{epsilon} and \texttt{epsilon-prepare} are installed into the sub-directory \texttt{bin} of the installation path.
      As before, \texttt{make install} might need \texttt{root} access depending on the installation prefix.
      For a non-standard installation prefix, the environment variable \texttt{PATH} should be adjusted to include the sub-directory \texttt{bin} of \texttt{/path/to/install} so that the programs \texttt{epsilon} and \texttt{epsilon-prepare} can be found by the shell.

      It is also possible to build \texttt{epsilon} and \texttt{epsilon-prepare} individually.
      This can be done by changing into the corresponding sub-directory and running \texttt{cmake} and \texttt{make} from within there.

      The \texttt{epsilon}-repository also offers a \texttt{Mathematica} package \texttt{EpsilonTools.m} found in the sub-directory \texttt{mma/}.
      Run
      \begin{lstlisting}[basicstyle=\ttfamily,xleftmargin=-4em]
        ./install.sh
      \end{lstlisting}
      from within this sub-directory to install \texttt{EpsilonTools.m} into the \texttt{Applications/} directory of your Mathematica installation.
    \subsection{Input/Output format} \label{sect:format}
      \texttt{epsilon} uses its own file format to represent a system of differential equations of the form \eqref{eq:M}, where every line represents one coefficient matrix.
      A line starts with either `\texttt{A[$x_j$,$k$]:}' or `\texttt{B[$k$]:}' followed by a matrix.
      The matrix is stored as a list of rows, where each row is itself a list of matrix elements.
      Lists are enclosed in curly-braces and list entries are separated by commas.
      A line starting with `\texttt{A[$x_j$,$k$]:}' (`\texttt{B[$k$]:}') represents a matrix $\M^{(x_j)}_k$ ($\M_k$) in \eqref{eq:M}.

      The name of the symbol used to represent $\epsilon$ is fixed to \texttt{ep}.

      An example of an input file for \texttt{epsilon} is given in section~\ref{sect:example}.
    \subsection{Usage of \texttt{epsilon-prepare}}
      The tool \texttt{epsilon-prepare} is used to convert a matrix in \texttt{Mathematica} format (a list of lists) into \texttt{epsilon} input format (see section \ref{sect:format}).
      If, for example, the file containing a matrix in \texttt{Mathematica} format is called \texttt{matrix.m}, then the command
      \begin{lstlisting}[basicstyle=\ttfamily,xleftmargin=-4em]
        epsilon-prepare matrix.m > matrix.dat
      \end{lstlisting}
      is used to create a file \texttt{matrix.dat} in \texttt{epsilon} format.
      The matrix elements of the input matrix are expected to be rational functions in $\epsilon$ and $x$ (represented by the symbols \texttt{ep} and \texttt{x}, respectively).
      The command \texttt{epsilon-prepare} performs a partial fraction decomposition over complex numbers in the variable $x$ of the matrix elements and results in an expression of the form of \eqref{eq:M}.
      
      To perform a partial fractioning of a rational function, the zeros of its denominator have to be determined.
      Therefore, \texttt{epsilon-prepare} applies \texttt{GiNaC}'s polynomial factorization algorithm to the denominator in order to factorize it into polynomials that are irreducible over the integers.
      In a second step, the zeros of all factors are found individually.

      This step will fail, if the considered irreducible polynomial has a degree larger than two.
      An error is thrown as well if the system is not in form \eqref{eq:M}, i.e. a zero depends on the parameter $\epsilon$.

      The additional symbols listed in table \ref{tbl:prep_syms} might be introduced by \texttt{epsilon-prepare}.
      See section \ref{sect:usage_field} for how to make \texttt{epsilon} accept them.

      \begin{table}
        \centering
        \begin{tabular}{ccc}
          symbol & polymod & expression \\ \hline
          \texttt{i} & $\texttt{i}^2 + 1$ & $i$ \\
          \texttt{r$N$} & $\texttt{r$N$}^2 - \texttt{r$N$} + (1+N)/4$ & $\frac{1+i\sqrt{N}}2$ \\
          \texttt{q$N$} & $\texttt{q$N$}^2 - \texttt{q$N$} + (1-N)/4$ & $\frac{1+\sqrt{N}}2$ \\
          \texttt{sqrt$N$} & $\texttt{sqrt$N$}^2 - N$ & $\sqrt{N}$ \\
          \texttt{isqrt$N$} & $\texttt{isqrt$N$}^2 + N$ & $i\sqrt{N}$
        \end{tabular}
        \caption{
          \label{tbl:prep_syms}
          Additional symbols used by \texttt{epsilon-prepare} to represent zeros of quadratic polynomials.
          $N$ is a positive integer.
        }          
      \end{table}
    \subsection{Usage of \texttt{epsilon}}
      The general command syntax for the tool \texttt{epsilon} is
      \begin{lstlisting}[basicstyle=\ttfamily,xleftmargin=-4em]
        epsilon [OPTIONS] JOBS...
      \end{lstlisting}
      The path to the \texttt{Fermat} binary can be set inside the environment variable \texttt{FERMAT}.
      If this variable is not set explicitly, \texttt{epsilon} will look for a binary \texttt{fer64} inside the directories defined in the environment variable \texttt{PATH}.
      Options are set once at the start of \texttt{epsilon}.
      Jobs are processed one by one in the same order they are defined on the command line.
      Some jobs will perform transformations to the system.
      These transformations are stored in a so-called internal transformation queue in RAM and can also be written to an external file.
      \subsubsection*{Options:}
        \begin{itemize}
          \item
            \uoption{--verbose}{Enable verbose output.}

            This option prints out all communication between \texttt{libFermat} and \texttt{Fermat} during a regular run.
            This is useful as a debug tool.
          \item
            \uoption{--timings}{Enable timings.}
            
            This option prints the elapsed real time after every job and the total time of the complete run.
          \item
            \uoption{--symbols \textit{symbols}}{Adjoin additional symbols to \texttt{Fermat}.}

            This option adds the specified symbols to \texttt{Fermat}.
            The symbols defined in this option are the very first variables adjoint to \texttt{Fermat} followed by the symbol \texttt{ep} and further internally used symbols.
            \texttt{\textit{symbols}} has to be a comma-separated list.
          \item
            \uoption{--echelon-fermat}{Use the \texttt{Redrowech} function in \texttt{Fermat} to solve linear systems.}

            At various points in the code linear systems of equations have to be solved.
            Use this option to choose the \texttt{Redrowech} function over our own implementation of Gaussian elimination.
        \end{itemize}
      \subsubsection*{Jobs:}
        \begin{itemize}
          \item
            \uoption{--fermat \textit{file}}{Execute \texttt{Fermat} commands.}

            This job reads \texttt{\textit{file}} line by line and sends all non-empty lines to \texttt{Fermat}.
          \item
            \uoption{--load \textit{file} \textit{start} \textit{end}}{Load system.}

            This job loads \texttt{\textit{file}} and activates the block $\{\texttt{\textit{start}},\texttt{\textit{end}}\}$.
            Hereby \texttt{\textit{file}} is expected to be in the format specified in section~\ref{sect:format}
          \item
            \uoption{--write \textit{file}}{Write system.}

            This job writes the system of differential equations to \texttt{\textit{file}} using the format specified in section~\ref{sect:format}.
          \item
            \uoption{--queue \textit{file}}{Set external transformation queue.}

            This job enables an external transformation queue.
            An external transformation queue is a file containing all transformations already performed by \texttt{epsilon} during a run.
            This is particularly useful in connection with the options \texttt{--load-queue} and \texttt{--replay} to restore an aborted run to the state after the last successful transformation.
          \item
            \uoption{--load-queue \textit{filename}}{Load transformation queue.}

            This job loads an external transformation queue from \texttt{\textit{filename}} into an internal transformation queue stored in RAM.
            This job does \emph{not} apply the transformations stored in the file to the system.
          \item
            \uoption{--replay}{Apply internal transformation queue.}

            This job `replays' the internal transformation queue, i.e. the transformations in the queue are applied to the system one by one.
            It should only be used immediately after \texttt{--load-queue}.
          \item
            \uoption{--export \textit{file}}{Export transformation matrix.}

            This job computes a transformation matrix out of the transformations inside the internal transformation queue.
            The matrix is written in \texttt{Mathematica} format to \texttt{\textit{file}}.
          \item
            \uoption{--block \textit{start} \textit{end}}{Activate a block.}

            This job activates the block $\{\texttt{\textit{start}},\texttt{\textit{end}}\}$.
          \item
            \uoption{--fuchsify}{Transform active block into Fuchsian form.}

            This job reduces the active block to Fuchsian form \eqref{eq:fuchsian}.
            See section~\ref{sect:lee} for details.
          \item
            \uoption{--normalize}{Normalize eigenvalues.}

            This job normalizes the eigenvalues of all residue matrices making them proportional to $\epsilon$.
            See section~\ref{sect:lee} for details.
          \item
            \uoption{--factorep}{Factor out $\epsilon$ (auto detect $\mu$).}

            This job transforms the active block into $\epsilon$-form \eqref{eq:epform} using the method described in section~\ref{sect:lee}.
            The variable $\mu$ in \eqref{eq:epfact} is left as an unknown and will be fixed only \emph{after} the system is solved to ensure that the transformation is invertible.
          \item
            \uoption{--factorep-at \textit{mu}}{Factor out $\epsilon$ (with predefined $\mu$).}

            This job transforms the active block into $\epsilon$-form \eqref{eq:epform} using the method described in section~\ref{sect:lee}.
            In this variant, the variable $\mu$ in \eqref{eq:epfact} is set to \texttt{\textit{mu}} \emph{before} the system is solved.
            This is faster than \texttt{--factorep} because \texttt{epsilon} has to deal with one less variable in the polynomials.
            Unfortunately, an unlucky choice of \texttt{\textit{mu}} can hit a pole in the matrix elements of the system or one can end up with an uninvertible transformation.
            In both cases an error is thrown.
          \item
            \uoption{--left-fuchsify}{Fuchsify off-diagonal block.}
            
            This job is used to transform the block left of the active block (block $\B$ in \eqref{eq:triangular}) to Fuchsian form.
            See section~\ref{sect:lee} for details.
          \item
            \uoption{--dyson \textit{file} \textit{order} \textit{type} \textit{format}}{Generate Dyson operator.}

            This job writes a Dyson operator $U(x,x_0)$ for the active block up to order \texttt{\textit{order}} in $\epsilon$ to \texttt{\textit{file}}.
            The active block has to be in $\epsilon$-form.
            The Dyson operator fulfills
            \[
              \frac{\partial}{\partial x}
              U(x,x_0)
              =
              \epsilon \sum\limits_{x_j \in S} 
              \frac{\widehat\C^{(x_j)}_0}{x-x_j}
              U(x,x_0)
              \,, \quad
              U(x_0,x_0) = \id\,.
            \]
            The option \texttt{\textit{type}} specifies the type of multiple polylogarithms in the output and can be set to \texttt{GPL}, \texttt{HPL} or \texttt{HPLalt} for Goncharov polylogarithms\cite{Goncharov:2010jf} or harmonic polylogarithms in the ``a''- or ``m''-notation\cite{Remiddi:1999ew,Maitre:2005uu}, respectively.
            \texttt{\textit{format}} should be \texttt{mma} or \texttt{form} to specify \texttt{Mathematica} or \texttt{FORM}\cite{Vermaseren:2000nd} as output format.
        \end{itemize}
    \subsection{Using field extensions} \label{sect:usage_field}
        The \texttt{Fermat} computer algebra system works with multivariate polynomials over the ground ring $\mathds{Z}$ and the corresponding quotient field, the rational functions over $\mathds{Z}$.
        Fortunately, \texttt{Fermat} offers a way to extend the ground ring by setting so-called polymods.

        A polymod $p(\xi)$ is a univariate polynomial in $\xi$, where $\xi$ is one of the variables adjoint to \texttt{Fermat}.
        This instates a new quotient ring $\mathds{Z}[\xi]/\langle p(\xi) \rangle$ as the ground ring, forcing the condition $p(\xi) = 0$ onto the variable $\xi$.
        In other words every polynomial $q(\xi) \in \mathds{Z}[\xi]$ encountered by \texttt{Fermat} is replaced immediately by the remainder of a polynomial division $q(\xi)\div p(\xi)$.
        For more details see the \texttt{Fermat} manual\cite{Lewis:fermat}.

        The polymod $i^2 + 1$ for example leads to the quotient ring $\mathds{Z}[i]/\langle i^2 + 1\rangle$ which is equivalent to the Gaussian integers, a field extension of the integers by a number $i$ with $i^2 = -1$.
        Also other complex numbers can be represented in \texttt{Fermat} using polymods.
        The complex numbers introduced via \texttt{epsilon-prepare} can be represented in \texttt{Fermat} using the polymods listed in table~\ref{tbl:prep_syms}.
       
        The syntax to set a polymod in \texttt{Fermat} is \texttt{\&(P=\textit{polymod},1)}, where the variable of the polymod must be the first symbol adjoint to \texttt{Fermat} which does not have a polymod assigned yet.
        This \texttt{Fermat} command should be stored in a text file and can be read in by \texttt{epsilon} with the \texttt{--fermat} job.

        For internal reasons \texttt{Fermat} runs very slow if more than one polymod is assigned to it.
        Therefore, in complicated cases with more than one complex number, it is useful to set a polymod only for the most frequent variable appearing in the block to work with next.
        As the order of symbols once set cannot be changed inside \texttt{Fermat}, the only way to change the `active' polymod is by saving the system with \texttt{--write} and reload it into a new session after the symbols are adjoined in a different order.
        In our tests we were able to solve huge systems with up to three complex numbers applying this strategy.
    \subsection{Usage of \texttt{EpsilonTools.m}}
      The Mathematica package \texttt{EpsilonTools.m} provides functions which help to set up \texttt{epsilon} and to work with the \texttt{epsilon} input/output file format (see section~\ref{sect:format}).
      It is not essential in order to run \texttt{epsilon}.
      After a successful installation, the package can be loaded into a Mathematica session with
      \begin{lstlisting}[basicstyle=\ttfamily,xleftmargin=-4em]
        <<EpsilonTools`
      \end{lstlisting}
      \texttt{EpsilonTools.m} provides three functions:
      \begin{itemize}
        \item
          \texttt{EpsilonSymRules[\textit{expression}]}

          This function scans \texttt{\textit{expression}} for symbols of the form given in table~\ref{tbl:prep_syms} and compiles a list of rules for these symbols to their corresponding Mathematica expressions, e.g.
          \begin{lstlisting}[basicstyle=\ttfamily,xleftmargin=-4em]
            {r3->(1+I*Sqrt[3])/2, q5->(1+Sqrt[5])/2, i->I}
          \end{lstlisting}          
        \item
          \texttt{EpsilonRead[\textit{file}]}
    
          This function reads \texttt{\textit{file}} into Mathematica where \texttt{\textit{file}} is in the format described in section~\ref{sect:format}.
          
          Options:
          \begin{itemize}
            \item 
              \texttt{ReplaceSymbols} (default: \texttt{True})

              If this option is set to \texttt{True}, all symbols introduced by \texttt{epsilon-prepare} will be replaced by their corresponding Mathematica expression.
            \item
              \texttt{CheckFuchsian} (default: \texttt{False})

              If this option is set to \texttt{True}, the function returns \texttt{\$Failed} if the system in \texttt{\textit{file}} is not in Fuchsian form.
            \item
              \texttt{CheckEpsilon} (default: \texttt{False})
              
              If this option is set to \texttt{True}, the function returns \texttt{\$Failed} if the system in \texttt{\textit{file}} is not in $\epsilon$-form.
          \end{itemize}
        \item
          \texttt{EpsilonBlocks[\textit{M}]}

          This function scans for a block-triangular structure of the matrix \texttt{\textit{M}} and returns a list of the boundaries of all diagonal blocks.
          The returned boundaries are in the form \texttt{\{\textit{start},\textit{end}\}}.
          The values \texttt{\textit{start}} and \texttt{\textit{end}} can be used in the \texttt{--block} job of \texttt{epsilon}.
      \end{itemize}
  \section{A physical example} \label{sect:example}
    \begin{figure}
      \centering
      \begin{subfigure}[b]{.24\textwidth}
        \centering
        \includegraphics[scale=1.5]{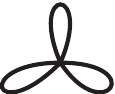}
        \caption*{$I_1$}
      \end{subfigure}%
      \begin{subfigure}[b]{.24\textwidth}
        \centering
        \includegraphics[scale=1.5]{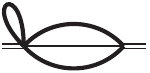}
        \caption*{$I_2$}
      \end{subfigure}%
      \begin{subfigure}[b]{.24\textwidth}
        \centering
        \includegraphics[scale=1.5]{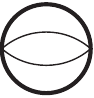}
        \caption*{$I_4$}
      \end{subfigure}%
      \begin{subfigure}[b]{.24\textwidth}
        \centering
        \includegraphics[scale=1.5]{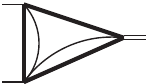}
        \caption*{$I_5$}
      \end{subfigure} \\
      \begin{subfigure}[b]{.24\textwidth}
        \centering
        \includegraphics[scale=1.5]{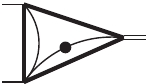}
        \caption*{$I_6$}
      \end{subfigure}%
      \begin{subfigure}[b]{.24\textwidth}
        \centering
        \includegraphics[scale=1.5]{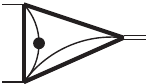}
        \caption*{$I_7$}
      \end{subfigure}%
      \begin{subfigure}[b]{.24\textwidth}
        \centering
        \includegraphics[scale=1.5]{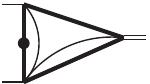}
        \caption*{$I_8$}
      \end{subfigure}%
      \begin{subfigure}[b]{.24\textwidth}
        \centering
        \includegraphics[scale=1.5]{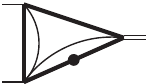}
        \caption*{$I_9$}
      \end{subfigure}
      \caption{ \label{fig:example}
        Three-loop master integrals to be solved with \texttt{epsilon}. 
        The not pictured integral $I_3$ has the same topology as $I_2$ but with an additional numerator.
        The thick (thin) lines are massive (massless).
        The thin external lines carry the momenta $q_1$ and $q_2$, while the double line carries the momentum $q_1+q_2$.
      }
    \end{figure}
    As an example, we consider a set of three-loop master integrals $\{I_j\}$ with $j=1,\dots,9$ in $d=4-2\epsilon$ dimensions with internal lines of mass one or zero.
    A graphical representation of the master integrals except $I_3$ is contained in fig.~\ref{fig:example}.
    Further we define $I_3$ as
    \[
      I_3 
      =
        \int d^dp
        \int d^dl
        \int d^dk
        \;
        \frac{
          (k-l)^2
        }{
          [(p+q_1)^2 - 1]
          [l^2-1]
          [(p-l+k)^2 - 1]
          [(l-p+q_2)^2]
        }\,.          
    \]
    The kinematics is given by
    \[
      q_1^2 = q_2^2 = 0\,, \quad
      q_1\cdot q_2 = -\frac{(1-x)^2}{2x}\,.
    \]      
    The vector $\vec{f} = (I_1,\dots,I_9)$ obeys a differential equation
    \[
      \frac{\partial}{\partial x} \vec{f} = \M(x,\epsilon) \vec{f}\,,
    \]
    where the $9\times9$-matrix $\M(x,\epsilon)$ has the structure
    \begin{equation} \label{eq:ex_matrix}
      \M(x,\epsilon) = \left( 
      \begin{BMAT}{ccccccccc}{ccccccccc}
        0 & 0 & 0 & 0 & 0 & 0 & 0 & 0 & 0 \\
        \ast & \ast & \ast & 0 & 0 & 0 & 0 & 0 & 0 \\
        \ast & \ast & \ast & 0 & 0 & 0 & 0 & 0 & 0 \\
        0 & 0 & 0 & 0 & 0 & 0 & 0 & 0 & 0 \\
        \ast & \ast & \ast & \ast & \ast & \ast & \ast & \ast & \ast \\
        \ast & \ast & \ast & \ast & \ast & \ast & \ast & \ast & \ast \\
        \ast & \ast & \ast & \ast & \ast & \ast & \ast & \ast & \ast \\
        \ast & \ast & \ast & \ast & \ast & \ast & \ast & \ast & \ast \\
        \ast & \ast & \ast & \ast & \ast & \ast & \ast & \ast & \ast 
        \addpath{(1,9,|)drrddrdrrrrr}
      \end{BMAT} \right)\,.          
    \end{equation}
    The $\ast$ represents any non-zero entry and we point out the block-triangular structure of the system.
    First, this matrix should be stored in a \texttt{Mathematica} compatible file \texttt{matrix.m}.
    The next step is to convert this file to the format described in section~\ref{sect:format} via
    \begin{lstlisting}[basicstyle=\ttfamily,xleftmargin=-2em]
      epsilon-prepare matrix.m > matrix.dat
    \end{lstlisting}
    The generated file \texttt{matrix.dat} should read
    \begin{lstlisting}[basicstyle=\ttfamily,xleftmargin=-2em]
      A[r3,0]:  	{{0,0,0,0,0,0,0,0,0},{0,0,0,0,0,0,0,0 ...
      A[-1,0]:  	{{0,0,0,0,0,0,0,0,0},{1-ep,5-6*ep,-6+ ...
      A[-1,1]:  	{{0,0,0,0,0,0,0,0,0},{0,0,0,0,0,0,0,0 ...
      A[-1,2]:  	{{0,0,0,0,0,0,0,0,0},{0,0,0,0,0,0,0,0 ...
      A[1-r3,0]:	{{0,0,0,0,0,0,0,0,0},{0,0,0,0,0,0,0,0 ...
      A[1,0]:   	{{0,0,0,0,0,0,0,0,0},{-1+ep,-11+10*ep ...
      A[1,1]:   	{{0,0,0,0,0,0,0,0,0},{0,0,0,0,0,0,0,0 ...
      A[1,2]:   	{{0,0,0,0,0,0,0,0,0},{0,0,0,0,0,0,0,0 ...
      A[1,3]:   	{{0,0,0,0,0,0,0,0,0},{0,0,0,0,0,0,0,0 ...
      A[1,4]:   	{{0,0,0,0,0,0,0,0,0},{0,0,0,0,0,0,0,0 ...
      A[0,0]:   	{{0,0,0,0,0,0,0,0,0},{0,3-2*ep,0,0,0, ...
      A[0,1]:   	{{0,0,0,0,0,0,0,0,0},{0,0,0,0,0,0,0,0 ...
      B[0]:     	{{0,0,0,0,0,0,0,0,0},{0,0,0,0,0,0,0,0 ...
    \end{lstlisting}      
    The arguments of \texttt{A} and \texttt{B} reveal singularities at $\{r_3,-1,1-r_3,1,0,\infty\}$ with Poincar\'{e} ranks $\{0,2,0,4,1,1\}$, respectively.
    The symbol $r_3$ introduced by \texttt{epsilon-prepare} is a root of the polynomial $r_3^2 - r_3 + 1$ and is given by $r_3 = (1+i\sqrt3)/2$ (see table~\ref{tbl:prep_syms}).
    Hence, we need a file \texttt{enable.r3.fer} to set a polynomial for \texttt{Fermat} to mod out with containing
    \begin{lstlisting}[basicstyle=\ttfamily,xleftmargin=-2em]
      &(P=r3^2-r3+1,1)
    \end{lstlisting}      
    To understand the origin of $r_3$ we consider for example the partial fraction decomposition over the complex of $[\M(x,\epsilon)]_{51}$:
    \begin{align*}
      [\M(x,\epsilon)]_{51}
      &= 
      \frac{(\epsilon-1)^2(1+x)\left(1+4x+x^2+\epsilon\left(1-10x+x^2\right)\right)}{4\epsilon(2\epsilon-1)(x-1)^3\left(x^2-x+1\right)} \\
      &= 
        \frac1{x-r_3} \left\{
          \frac{5-9\epsilon^3+23\epsilon^2-19\epsilon}{4(2\epsilon^2-\epsilon)}
        \right\}
        +
        \frac1{x-(1-r_3)} \left\{
          \frac{5-9\epsilon^3+23\epsilon^2-19\epsilon}{4(2\epsilon^2-\epsilon)}
        \right\} \\ &\quad
        +
        \frac1{x-1} \left\{
          \frac{-5+9\epsilon^3-23\epsilon^2+19\epsilon}{2(2\epsilon^2-\epsilon)}
        \right\}
        +
        \frac1{(x-1)^2} \left\{
           \frac{3-4\epsilon^3+11\epsilon^2-10\epsilon}{2(2\epsilon^2-\epsilon)}
        \right\} \\ &\quad
        + 
        \frac1{(x-1)^3} \left\{
          \frac{3-4\epsilon^3+11\epsilon^2-10\epsilon}{2\epsilon^2-\epsilon}
        \right\}\,.
    \end{align*}

    The possible blocks to run \texttt{epsilon} with can be either read off from the matrix \eqref{eq:ex_matrix} or determined by the function \texttt{EpsilonBlocks} of the \texttt{Mathematica} package \texttt{EpsilonTools.m}.
    In this case, the possible blocks are
    \[
      \{1,1\},
      \{2,3\},
      \{4,4\},
      \{5,9\}\,.
    \]
    Now we can run \texttt{epsilon} with the command
    \begin{lstlisting}[basicstyle=\ttfamily,xleftmargin=-2em]
      epsilon --timings --symbols r3 --load matrix.dat 1 1 \
        --queue out.queue \
        --fermat enable.r3.fer \
        --fuchsify --normalize --factorep-at -1 \
        --block 2 3 \
        --fuchsify --normalize --factorep-at -1 --left-fuchsify \
        --block 4 4 \
        --fuchsify --normalize --factorep-at -1 --left-fuchsify \
        --block 5 9 \
        --fuchsify --normalize --factorep-at  1 --left-fuchsify \
        --block 1 9 \
        --factorep-at -1 \
        --write epsilon.dat --export transformation.m
    \end{lstlisting}
    The \texttt{--load} job in the first line loads the file \texttt{matrix.dat} with an active block $\{1,1\}$; the \texttt{--block} jobs are used to change the active block.
    In all blocks except for block $\{5,9\}$ we factor out $\epsilon$ at $\mu=-1$.
    For this block $\mu=-1$ would lead to a singular system so we choose $\mu=1$ instead.
    The final system is written into the file \texttt{epsilon.dat} and the corresponding transformation matrix into \texttt{transformation.m}.
    The whole reduction process takes about three minutes on an Intel Core i5-3320M CPU with 2.60 GHz.
    The package \texttt{EpsilonTools.m} offers the function \texttt{EpsilonRead} for reading the file \texttt{epsilon.dat} into \texttt{Mathematica}.
    The result is in $\epsilon$-form \eqref{eq:epform}, with
    \[  \widehat\M^{(-1)}_0 = {\tiny \left( \begin{BMAT}{ccccccccc}{ccccccccc}
        0 & 0 & 0 & 0 & 0 & 0 & 0 & 0 & 0 \\
        \frac{1}{5} & -18 & \frac{36}{5} & 0 & 0 & 0 & 0 & 0 & 0 \\
        \frac{1}{3} & -30 & 12 & 0 & 0 & 0 & 0 & 0 & 0 \\
        0 & 0 & 0 & 0 & 0 & 0 & 0 & 0 & 0 \\
        \frac{7790185}{4743936} & \frac{-406887355}{3162624} & \frac{13303239}{263552} & \frac{-3715319}{6325248} & \frac{-21}{58} & \frac{323}{58} & 1 & \frac{-121}{58} & \frac{263}{116} \\
        \frac{222799291}{75902976} & \frac{-557897095}{12650496} & \frac{13303239}{2108416} & \frac{-85452337}{12650496} & \frac{-483}{116} & \frac{7429}{116} & \frac{23}{2} & \frac{-2783}{116} & \frac{6049}{232} \\
        \frac{-19366519759}{2201186304} & \frac{71280791995}{366864384} & \frac{-2870982795}{61144064} & \frac{6728442709}{366864384} & \frac{38031}{3364} & \frac{-584953}{3364} & \frac{-1811}{58} & \frac{219131}{3364} & \frac{-476293}{6728} \\
        \frac{90725693}{37951488} & \frac{2996225}{89088} & \frac{-27685119}{1054208} & \frac{-48299147}{6325248} & \frac{-273}{58} & \frac{4199}{58} & 13 & \frac{-1573}{58} & \frac{3419}{116} \\
        \frac{-48778543}{37951488} & \frac{126440695}{6325248} & \frac{-3235923}{1054208} & \frac{18576595}{6325248} & \frac{105}{58} & \frac{-1615}{58} & -5 & \frac{605}{58} & \frac{-1315}{116} 
      \end{BMAT} \right)}\,,
    \]
    \[  \widehat\M^{(0)}_0 = {\tiny \left( \begin{BMAT}{ccccccccc}{ccccccccc}
        0 & 0 & 0 & 0 & 0 & 0 & 0 & 0 & 0 \\
        \frac{-2}{5} & 26 & -12 & 0 & 0 & 0 & 0 & 0 & 0 \\
        \frac{-7}{9} & 50 & -23 & 0 & 0 & 0 & 0 & 0 & 0 \\
        0 & 0 & 0 & 0 & 0 & 0 & 0 & 0 & 0 \\
        \frac{-30321797}{7115904} & \frac{356550775}{2371968} & \frac{-119849}{2059} & \frac{19535387}{4743936} & \frac{11137}{5568} & \frac{-51841}{1392} & \frac{-1225}{192} & \frac{35135}{2784} & \frac{-189319}{11136} \\
        \frac{-686135525}{113854464} & \frac{223518385}{37951488} & \frac{29123307}{1054208} & \frac{529852429}{37951488} & \frac{419699}{44544} & \frac{-1585715}{11136} & \frac{-37835}{1536} & \frac{1147213}{22272} & \frac{-4875557}{89088} \\
        \frac{49678009745}{3301779456} & \frac{-229880569165}{1100593152} & \frac{1437349057}{30572032} & \frac{-30031402873}{1100593152} & \frac{-25717127}{1291776} & \frac{91504967}{322944} & \frac{2219903}{44544} & \frac{-65705785}{645888} & \frac{274163633}{2583552} \\
        \frac{-323472451}{56927232} & \frac{-2193835945}{18975744} & \frac{50456429}{527104} & \frac{344565875}{18975744} & \frac{262093}{22272} & \frac{-1020493}{5568} & \frac{-24181}{768} & \frac{746867}{11136} & \frac{-3104155}{44544} \\
        \frac{158320529}{56927232} & \frac{402093395}{18975744} & \frac{-14262031}{527104} & \frac{-137227105}{18975744} & \frac{-101759}{22272} & \frac{404543}{5568} & \frac{9527}{768} & \frac{-290113}{11136} & \frac{1313657}{44544} 
      \end{BMAT} \right)}\,,
    \]
    \[  \widehat\M^{(1)}_0 = {\tiny \left( \begin{BMAT}{ccccccccc}{ccccccccc}
        0 & 0 & 0 & 0 & 0 & 0 & 0 & 0 & 0 \\
        \frac{1}{5} & -10 & \frac{24}{5} & 0 & 0 & 0 & 0 & 0 & 0 \\
        \frac{4}{9} & -20 & 10 & 0 & 0 & 0 & 0 & 0 & 0 \\
        0 & 0 & 0 & 0 & 0 & 0 & 0 & 0 & 0 \\
        \frac{41108207}{14231808} & \frac{-636997435}{9487872} & \frac{21452971}{790656} & \frac{28883609}{18975744} & \frac{673}{116} & \frac{-1484}{87} & \frac{-13}{4} & \frac{1651}{174} & \frac{1785}{232} \\
        \frac{733595729}{227708928} & \frac{22172065}{4743936} & \frac{-1318339}{109056} & \frac{-334258861}{75902976} & \frac{-1507}{464} & \frac{5221}{87} & \frac{167}{16} & \frac{-14597}{696} & \frac{24933}{928} \\
        \frac{-41044567181}{6603558912} & \frac{2404770185}{137574144} & \frac{-2068713589}{91716096} & \frac{20579631337}{2201186304} & \frac{194887}{13456} & \frac{-331405}{2523} & \frac{-11707}{464} & \frac{952145}{20184} & \frac{-1183713}{26912} \\
        \frac{391067287}{113854464} & \frac{50935825}{1185984} & \frac{-71070457}{1581312} & \frac{-281285603}{37951488} & \frac{-1053}{232} & \frac{7870}{87} & \frac{121}{8} & \frac{-11539}{348} & \frac{17115}{464} \\
        \frac{-188522477}{113854464} & \frac{-28164515}{4743936} & \frac{18097199}{1581312} & \frac{43744885}{37951488} & \frac{-237}{232} & \frac{-1634}{87} & \frac{-23}{8} & \frac{1877}{348} & \frac{-7157}{464} 
      \end{BMAT} \right)}\,,
    \]
    \[  \widehat\M^{(r_3)}_0 = {\tiny \left( \begin{BMAT}{ccccccccc}{ccccccccc}
        0 & 0 & 0 & 0 & 0 & 0 & 0 & 0 & 0 \\
        0 & 0 & 0 & 0 & 0 & 0 & 0 & 0 & 0 \\
        0 & 0 & 0 & 0 & 0 & 0 & 0 & 0 & 0 \\
        0 & 0 & 0 & 0 & 0 & 0 & 0 & 0 & 0 \\
        \frac{-119849}{889488} & \frac{1797735}{65888} & \frac{-838943}{49416} & \frac{-2996225}{1185984} & \frac{-2425}{696} & \frac{1355}{58} & \frac{97}{24} & \frac{-1053}{116} & \frac{5839}{1392} \\
        \frac{-3715319}{56927232} & \frac{55729785}{4216832} & \frac{-26007233}{3162624} & \frac{-92882975}{75902976} & \frac{-75175}{44544} & \frac{42005}{3712} & \frac{3007}{1536} & \frac{-32643}{7424} & \frac{181009}{89088} \\
        \frac{-26486629}{1650889728} & \frac{397299435}{122288128} & \frac{-185406403}{91716096} & \frac{-662165725}{2201186304} & \frac{-535925}{1291776} & \frac{299455}{107648} & \frac{21437}{44544} & \frac{-232713}{215296} & \frac{1290419}{2583552} \\
        \frac{-2037433}{28463616} & \frac{30561495}{2108416} & \frac{-14262031}{1581312} & \frac{-50935825}{37951488} & \frac{-41225}{22272} & \frac{23035}{1856} & \frac{1649}{768} & \frac{-17901}{3712} & \frac{99263}{44544} \\
        \frac{2277131}{28463616} & \frac{-34156965}{2108416} & \frac{15939917}{1581312} & \frac{56928275}{37951488} & \frac{46075}{22272} & \frac{-25745}{1856} & \frac{-1843}{768} & \frac{20007}{3712} & \frac{-110941}{44544} 
      \end{BMAT} \right)}\,,
    \]
    \[  \widehat\M^{(1 - r_3)}_0 = {\tiny \left( \begin{BMAT}{ccccccccc}{ccccccccc}
        0 & 0 & 0 & 0 & 0 & 0 & 0 & 0 & 0 \\
        0 & 0 & 0 & 0 & 0 & 0 & 0 & 0 & 0 \\
        0 & 0 & 0 & 0 & 0 & 0 & 0 & 0 & 0 \\
        0 & 0 & 0 & 0 & 0 & 0 & 0 & 0 & 0 \\
        \frac{-119849}{889488} & \frac{1797735}{65888} & \frac{-838943}{49416} & \frac{-2996225}{1185984} & \frac{-2425}{696} & \frac{1355}{58} & \frac{97}{24} & \frac{-1053}{116} & \frac{5839}{1392} \\
        \frac{-3715319}{56927232} & \frac{55729785}{4216832} & \frac{-26007233}{3162624} & \frac{-92882975}{75902976} & \frac{-75175}{44544} & \frac{42005}{3712} & \frac{3007}{1536} & \frac{-32643}{7424} & \frac{181009}{89088} \\
        \frac{-26486629}{1650889728} & \frac{397299435}{122288128} & \frac{-185406403}{91716096} & \frac{-662165725}{2201186304} & \frac{-535925}{1291776} & \frac{299455}{107648} & \frac{21437}{44544} & \frac{-232713}{215296} & \frac{1290419}{2583552} \\
        \frac{-2037433}{28463616} & \frac{30561495}{2108416} & \frac{-14262031}{1581312} & \frac{-50935825}{37951488} & \frac{-41225}{22272} & \frac{23035}{1856} & \frac{1649}{768} & \frac{-17901}{3712} & \frac{99263}{44544} \\
        \frac{2277131}{28463616} & \frac{-34156965}{2108416} & \frac{15939917}{1581312} & \frac{56928275}{37951488} & \frac{46075}{22272} & \frac{-25745}{1856} & \frac{-1843}{768} & \frac{20007}{3712} & \frac{-110941}{44544} 
      \end{BMAT} \right)}\,.
    \]
      
  \section{Summary}
    In this paper we presented \texttt{epsilon}.
    The tool \texttt{epsilon} is an efficient implementation of an algorithm proposed by R.N.~Lee to reduce a system of ordinary differential equations with rational coefficients to a canonical form, where the right hand side is proportional to $\epsilon$. 

    In physically relevant situations, the small parameter $\epsilon$ usually is a regulator in dimensional regularization (e.g. in $d=4-2\epsilon$ dimensions).
    We showed its applicability in a three-loop example and demonstrated the possibility to reduce systems with complex singular points.

  \section*{Acknowledgments}
    The author wants to thank Robert Harlander for useful discussions and comments on the manuscript.
    This work was supported by BMBF contract 05H15PACC1.
    The computing resources were granted by RWTH Aachen University under project rwth0119.
    The Feynman diagrams in this article have been drawn with \texttt{JaxoDraw}~\cite{Binosi:2008ig} based on \texttt{Axodraw}~\cite{Vermaseren:1994je}.
  \appendix
  \section{Transformations} \label{sect:app_bal}
    \subsection{Balances}
      The main building blocks of Lee's algorithm are balances.
      In this section we describe how they act on a system in the form \eqref{eq:M} which is also assumed to be in block-triangular form \eqref{eq:triangular}.

      The projector $\P$ is determined by its impact on the active block $\C$.
      Hence, it has to be of the form
      \[
        \P = \begin{pmatrix}
          0 & 0 & 0 \\
          0 & \Q & 0 \\
          0 & 0 & 0
        \end{pmatrix}\,.
      \]        
      Naturally, this also modifies the blocks $\B$ and $\E$.
      In this appendix we omit writing down the $\epsilon$-dependencies explicitly.

      First, we consider a balance between two singularities $x_1, x_2 < \infty$ \eqref{eq:bal_x1_x2}:
      \[
        \T = {\cal B}(\P,x_1,x_2)\,, \quad
        \T^{-1} = {\cal B}(\P,x_2,x_1)\,.
      \]
      The transformation \eqref{eq:trans} can now be written in terms of the coefficient matrices.

      We find for the active block
      \begin{subequations}
        \begin{align}
          \begin{split} \label{eq:bal_x1_x2_Cx10}
            \widetilde \C^{(x_1)}_0 &=
            \C^{(x_1)}_0
            -
            \sum\limits_{n\geq0} \frac1{(x_2-x_1)^n} 
            \Q \C^{(x_1)}_n \Qbar 
            +
            \sum\limits_{x_j\in S\backslash\{x_1\}} 
            \sum\limits_{n\geq0} 
            \frac{x_1-x_2}{(x_1-x_j)^{n+1}} 
            \Qbar \C^{(x_j)}_n \Q \\ &\quad
            +  (x_1-x_2) \sum\limits_{n\geq0} x_1^n \Qbar \C_n \Q 
            + \Q\,, 
          \end{split} \\   
          \begin{split}
            \widetilde \C^{(x_1)}_{k>0} &=
            \C^{(x_1)}_k
            +
            (x_1-x_2) \Qbar \C^{(x_1)}_{k-1} \Q
            -
            \sum\limits_{n\geq0} \frac1{(x_2-x_1)^n} 
            \Q \C^{(x_1)}_{n+k} \Qbar\,,
          \end{split} \label{eq:bal_x1_x2_Cx1k} \\
          \begin{split}
            \widetilde \C^{(x_2)}_0 &=
            \C^{(x_2)}_0
            - 
            \sum\limits_{n\geq0} 
            \frac1{(x_1-x_2)^n} 
            \Qbar \C^{(x_2)}_n \Q
            +
            \sum\limits_{x_j\in S\backslash\{x_2\}}
            \sum\limits_{n\geq0} \frac{x_2-x_1}{(x_2-x_j)^{n+1}} 
            \Q \C^{(x_j)}_n \Qbar \\ &\quad
            + (x_2-x_1) \sum\limits_{n\geq0} x_2^n \Q \C_n \Qbar
            - \Q\,, 
          \end{split} \\
          \begin{split} 
            \widetilde \C^{(x_2)}_{k>0} &=
            \C^{(x_2)}_k
            +
            (x_2-x_1)
            \Q \C^{(x_2)}_{k-1} \Qbar
            - 
            \sum\limits_{n\geq0} 
            \frac1{(x_1-x_2)^n} 
            \Qbar \C^{(x_2)}_{n+k} \Q\,,
          \end{split} \label{eq:bal_x1_x2_Cx2k} \\
          \begin{split}
            \widetilde \C^{(x_j\neq x_1,x_2)}_k &=
            \C^{(x_j)}_k
            +
            \sum\limits_{n\geq0} 
            \frac{x_2-x_1}{(x_1-x_j)^{n+1}} 
            \Qbar \C^{(x_j)}_{n+k} \Q 
            + 
            \sum\limits_{n\geq0} 
            \frac{x_1-x_2}{(x_2-x_j)^{n+1}} 
            \Q \C^{(x_j)}_{n+k} \Qbar\,, 
          \end{split} \\
          \begin{split}
            \widetilde \C_k &= 
            \C_k +
            (x_1 - x_2)
            \sum\limits_{n\geq0} \left\{
              x_1^n \Qbar \C_{k+n+1} \Q
              - x_2^n \Q \C_{k+n+1} \Qbar
            \right\}\,,
          \end{split}            
        \end{align}
      \end{subequations}        
      and for the off-diagonal blocks
      \begin{align*}
        \widetilde \B^{(x_2)}_0 &=
        \B^{(x_2)}_0
        +
        \sum\limits_{x_j\in S\backslash\{x_2\}}
        \sum\limits_{n\geq0} \frac{x_2-x_1}{(x_2-x_j)^{n+1}} 
        \Q \B^{(x_j)}_n
        + (x_2-x_1) \sum\limits_{n\geq0} x_2^n \Q \B_n\,,
        \\
        \widetilde \B^{(x_2)}_{k>0} &=
        \B^{(x_2)}_k
        +
        (x_2-x_1)
        \Q \B^{(x_2)}_{k-1}\,,
        \\
        \widetilde \B^{(x_j\neq x_2)}_k &=
        \B^{(x_j)}_k
        + 
        \sum\limits_{n\geq0} 
        \frac{x_1-x_2}{(x_2-x_j)^{n+1}} 
        \Q \B^{(x_j)}_{n+k}\,, 
        \\
        \widetilde \B_k &= 
        \B_k +
        (x_2-x_1)
        \sum\limits_{n\geq0}
        x_2^n \Q \B_{k+n+1} \,,
        \\
        \widetilde \E^{(x_1)}_0 &=
        \E^{(x_1)}_0
        +
        \sum\limits_{x_j\in S\backslash\{x_1\}} 
        \sum\limits_{n\geq0} 
        \frac{x_1-x_2}{(x_1-x_j)^{n+1}} 
        \E^{(x_j)}_n \Q
        + (x_1-x_2) \sum\limits_{n\geq0} x_1^n \E_n \Q\,,
        \\
        \widetilde \E^{(x_1)}_{k>0} &=
        \E^{(x_1)}_k
        +
        (x_1-x_2) \E^{(x_1)}_{k-1} \Q\,,
        \\
        \widetilde \E^{(x_j\neq x_1)}_k &=
        \E^{(x_j)}_k
        +
        \sum\limits_{n\geq0} 
        \frac{x_2-x_1}{(x_1-x_j)^{n+1}} 
        \E^{(x_j)}_{n+k} \Q 
        \\
        \widetilde \E_k &= 
        \E_k +
        (x_1 - x_2)
        \sum\limits_{n\geq0}
        x_1^n \E_{k+n+1} \Q\,.
      \end{align*}
      All other blocks are unaffected.

      Next, we consider the case $x_1<\infty$, $x_2 = \infty$, i.e.
      \[
        \T = {\cal B}(\P,x_1,\infty)\,, \quad
        \T^{-1} = {\cal B}(\P,\infty,x_1)\,.
      \]
      In that case, the balances are given by \eqref{eq:bal_x1_inf} and \eqref{eq:bal_inf_x2}.
      Here the active block transforms as
      \begin{align*}
        \widetilde \C^{(x_1)}_0 &=
          \C^{(x_1)}_0 - \Qbar \C^{(x_1)}_0 \Q - \Q \C^{(x_1)}_0 \Qbar
          + \Q \C^{(x_1)}_1 \Qbar \\ &\quad
          + \sum\limits_{x_j \in S\backslash\{x_1\}}
            \sum\limits_{n=0}^\infty \frac1{(x_1-x_j)^{n+1}} \Qbar \C^{(x_j)}_n \Q
          + \sum\limits_{n=0}^\infty x_1^n \Qbar \C_n \Q
          + \Q\,,
        \\
        \widetilde \C^{(x_1)}_{k>0} &=
          \C^{(x_1)}_k - \Qbar \C^{(x_1)}_k \Q - \Q \C^{(x_1)}_k \Qbar
          + \Q \C^{(x_1)}_{k+1} \Qbar
          + \Qbar \C^{(x_1)}_{k-1} \Q\,,
        \\
        \widetilde \C^{(x_j\neq x_1)}_k &=
          \C^{(x_j)}_k - \Qbar \C^{(x_j)}_k \Q - \Q \C^{(x_j)}_k \Qbar
          + \Q \C^{(x_j)}_{k+1} \Qbar 
          + (x_j-x_1) \Q \C^{(x_j)}_k \Qbar \\ &\quad
          - 
          \sum\limits_{n=0}^\infty \frac1{(x_1-x_j)^{n+1}} \Qbar \C^{(x_j)}_{n+k} \Q \,,
        \\
        \widetilde \C_0 &=
          \C_0 - \Qbar \C_0 \Q - (1+x_1) \Q \C_0 \Qbar
          + \sum\limits_{x_j \in S} \Q \C^{(x_j)}_0 \Qbar 
          + \sum\limits_{n=0}^\infty x_1^n \Qbar \C_{n+1} \Q \,,
        \\
        \widetilde \C_{k>0} &=
          \C_k - \Qbar \C_k \Q - (1+x_1) \Q \C_k \Qbar
          + \Q \C_{k-1} \Qbar 
          + \sum\limits_{n=0}^\infty x_1^n \Qbar \C_{k+n+1} \Q \,,
      \end{align*}
      and the blocks $\B$ and $\E$ as
      \begin{align*}
        \widetilde \B^{(x_j)}_k &=
          \B^{(x_j)}_k - \Q \B^{(x_j)}_k 
          + \Q \B^{(x_j)}_{k+1}  
          + (x_j-x_1) \Q \B^{(x_j)}_k \,, 
        \\
        \widetilde \B_0 &=
          \B_0 - (1+x_1) \Q \B_0 
        + \sum\limits_{x_j \in S} \Q \B^{(x_j)}_0  \,,
        \\
        \widetilde \B_{k>0} &=
          \B_k - (1+x_1) \Q \B_k 
          + \Q \B_{k-1} \,,
        \\
        \widetilde \E^{(x_1)}_0 &=
          \E^{(x_1)}_0 - \E^{(x_1)}_0 \Q 
          + \sum\limits_{x_j \in S\backslash\{x_1\}}
            \sum\limits_{n\geq0} \frac1{(x_1-x_j)^{n+1}} \E^{(x_j)}_n \Q
          + \sum\limits_{n\geq0} x_1^n \E_n \Q\,,
        \\
        \widetilde \E^{(x_1)}_{k>0} &=
          \E^{(x_1)}_k - \E^{(x_1)}_k \Q 
          + \E^{(x_1)}_{k-1} \Q\,,
        \\
        \widetilde \E^{(x_j\neq x_1)}_k &=
          \E^{(x_j)}_k - \E^{(x_j)}_k \Q 
          - 
          \sum\limits_{n\geq0} \frac1{(x_1-x_j)^{n+1}} \E^{(x_j)}_{n+k} \Q \,,
        \\
        \widetilde \E_k &=
          \E_k - \E_k \Q 
          + \sum\limits_{n\geq0} x_1^n \E_{k+n+1} \Q \,.
      \end{align*}        

      Finally, we consider the case $x_1 = \infty$, $x_2 < \infty$, i.e.
      \[
        \T = {\cal B}(\P,\infty,x_2)\,, \quad
        \T^{-1} = {\cal B}(\P,x_2,\infty)\,.
      \]
      This is similar to the previous case.
      The active block transforms to
      \begin{align*}
        \widetilde \C^{(x_2)}_0 &=
          \C^{(x_2)}_0 - \Qbar \C^{(x_2)}_0 \Q - \Q \C^{(x_2)}_0 \Qbar
          +
          \Qbar \C^{(x_2)}_1 \Q \\ &\quad
          +
          \sum\limits_{x_j \in S \backslash\{x_2\}} 
          \sum\limits_{n\geq0} \frac1{(x_2-x_j)^{n+1}}
          \Q \C^{(x_j)}_n \Qbar 
          + 
          \sum\limits_{n\geq0} 
          x_2^n 
          \Q \C_n \Qbar
          -
          \Q\,,
        \\
        \widetilde \C^{(x_2)}_{k>0} &=
          \C^{(x_2)}_k - \Qbar \C^{(x_2)}_k \Q - \Q \C^{(x_2)}_k \Qbar
          +
          \Q \C^{(x_2)}_{k-1} \Qbar 
          +
          \Qbar \C^{(x_2)}_{k+1} \Q \,,
        \\
        \widetilde \C^{(x_j\neq x_2)}_k &=
          \C^{(x_j)}_k - \Qbar \C^{(x_j)}_k \Q - \Q \C^{(x_j)}_k \Qbar
          +
          \Qbar \C^{(x_j)}_{k+1} \Q 
          +
          (x_j-x_2)
          \Qbar \C^{(x_j)}_k \Q \\ &\quad 
          - 
          \sum\limits_{n\geq0} 
          \frac1{(x_2-x_j)^{n+1}} 
          \Q \C^{(x_j)}_{n+k} \Qbar\,, 
        \\
        \widetilde \C_0 &=
          \C_0 - \Qbar \C_0 \Q - \Q \C_0 \Qbar
          +
          \sum\limits_{n\geq0} 
          x_2^n 
          \Q \C_{n+1} \Qbar 
          - 
          x_2 
          \Qbar \C_0 \Q
          +
          \sum\limits_{x_j \in S} 
          \Qbar \C^{(x_j)}_0 \Q \,, \\
        \widetilde \C_{k>0} &=
          \C_k - \Qbar \C_k \Q - \Q \C_k \Qbar
          +
          \sum\limits_{n\geq0} 
          x_2^n 
          \Q \C_{k+n+1} \Qbar 
          +
          \Qbar \C_{k-1} \Q 
          - 
          x_2 
          \Qbar \C_k \Q \,,
      \end{align*}
      and the blocks $\B$ and $\E$ transform as
      \begin{align*}
        \widetilde \B^{(x_2)}_0 &=
          \B^{(x_2)}_0 - \Q \B^{(x_2)}_0
          +
          \sum\limits_{x_j \in S \backslash\{x_2\}} 
          \sum\limits_{n\geq0} \frac1{(x_2-x_j)^{n+1}}
          \Q \B^{(x_j)}_n  
          + 
          \sum\limits_{n\geq0} 
          x_2^n 
          \Q \B_n\,,
        \\
        \widetilde \B^{(x_2)}_{k>0} &=
          \B^{(x_2)}_k - \Q \B^{(x_2)}_k 
          +
          \Q \B^{(x_2)}_{k-1}\,, 
        \\
        \widetilde \B^{(x_j\neq x_2)}_k &=
          \B^{(x_j)}_k - \Q \B^{(x_j)}_k 
          - 
          \sum\limits_{n\geq0} 
          \frac1{(x_2-x_j)^{n+1}} 
          \Q \B^{(x_j)}_{n+k}\,, 
        \\
        \widetilde \B_k &=
          \B_k - \Q \B_k 
          +
          \sum\limits_{n\geq0} 
          x_2^n 
          \Q \B_{k+n+1} \,,
        \\
        \widetilde \E^{(x_j)}_k &=
          \E^{(x_j)}_k - \E^{(x_j)}_k \Q 
          +
          \E^{(x_j)}_{k+1} \Q 
          +
          (x_j-x_2)
          \E^{(x_j)}_k \Q \,,  
        \\
        \widetilde \E_0 &=
          \E_0 - (1+x_2) \E_0 \Q 
          +
          \sum\limits_{x_j \in S} 
          \E^{(x_j)}_0 \Q \,, \\
        \widetilde \E_{k>0} &=
          \E_k - (1+x_2)\E_k \Q 
          +
          \E_{k-1} \Q\,. 
      \end{align*}
    \subsection{Fuchsification of off-diagonal blocks}
      In this appendix we consider the transformation of the off-diagonal block $\B$ to Fuchsian form.
      We assume that the diagonal blocks $\A$ and $\C$ are already in $\epsilon$-form.
      The transformation needed has the form
      \begin{equation} \label{eq:lefttrans}
        \T = {\cal L}(x_1,k,\G)\,, \quad
        \T^{-1} = {\cal L}(x_1,k,-\G)\,,
      \end{equation}
      where we used the definitions \eqref{eq:offred}. 

      In addition to block $\B$ only block $\D$ is influenced by this transformation, i.e. the blocks $\A$ and $\C$ are unaffected.
      
      The transformation \eqref{eq:trans}, with $\T$ defined as in \eqref{eq:lefttrans} translates to rules for the coefficient matrices.
      For $x_1 < \infty$ we find
      \begin{subequations}
        \begin{align}
          \widetilde\B^{(x_1)}_k &= 
            \B^{(x_1)}_k
            + \C_0^{(x_1)}\widehat\G - \widehat\G\A_0^{(x_1)} 
            + k\widehat\G\,, \label{eq:offred_Bx1k}
          \\
          \widetilde\B^{(x_1)}_{n<k} &= 
            \B^{(x_1)}_n
            -\sum\limits_{x_j\in S \backslash\{x_1\}} 
            \frac{\C_0^{(x_j)}\widehat\G - \widehat\G\A_0^{(x_j)}}{(x_j-x_1)^{k-n}}\,,
          \\
          \widetilde\B^{(x_1)}_{n>k} &= 
            \B^{(x_1)}_n\,,
          \\
          \widetilde\B^{(x_j \neq x_1)}_0 &= 
            \B^{(x_j)}_0
            +
            \frac{\C_0^{(x_j)}\widehat\G - \widehat\G\A_0^{(x_j)}}{(x_j-x_1)^k}\,,
          \\
          \widetilde\B^{(x_j \neq x_1)}_{n>0} &= 
            \B^{(x_j)}_n\,,
          \\
          \widetilde\B_n &= 
            \B_n\,,
        \end{align}
      \end{subequations}        
      and
      \begin{align*}
        \widetilde\D^{(x_1)}_{n<k} &= 
          \D^{(x_1)}_n
          -
          \sum\limits_{x_j\in S\backslash\{x_1\}} 
          \sum\limits_{i\geq0}
          (-1)^i
          \frac{\binom{k+i-n-1}{i}}{(x_j-x_1)^{k+i-n}}
          \E^{(x_j)}_i\widehat\G \\ &\quad
          +
          \sum\limits_{i\geq0}
          x_1^i \binom{i+k-n-1}{i}
          \E_{i+k-n-1} \widehat\G\,,
          \\
        \widetilde\D^{(x_1)}_{n\geq k} &=
          \D^{(x_1)}_n
          + 
          \E^{(x_1)}_{n-k}\widehat\G\,, 
          \\
        \widetilde\D^{(x_{j\neq1})}_n &=
          \D^{(x_j)}_n
          +
          (-1)^k
          \sum\limits_{i\geq0}
          \frac{\binom{k+i-1}{i}}{(x_1-x_j)^{k+i}} 
          \E^{(x_j)}_{n+i}\widehat\G\,,
          \\
        \widetilde\D_n &=
          \D_n
          +
          \sum\limits_{m\geq0}
          (-1)^{m} 
          \sum\limits_{i\geq0}
          x_1^{m+i} 
          \binom{n+m}{n} 
          \binom{i+n+m+k}{i} 
          \E_{i+n+m+k} \widehat\G\,.
      \end{align*}
      The case $x_1=\infty$ leads to
      \begin{align*}
        \widetilde\B_{k-1} &=
          \B_{k-1}
          +
          \sum\limits_{x_j\in S} 
          \left[\C_0^{(x_j)}\G - \widehat\G\A^{(x_j)}\right] 
          -
          k\G\,,
          \\
        \widetilde\B_{n<k-1} &=
          \B_n
          +
          \sum\limits_{x_j\in S} 
          x_j^{k-n-1}
          \left[\C_0^{(x_j)}\widehat\G - \widehat\G\A_0^{(x_j)}\right] \,,
          \\
        \widetilde\B_{n>k-1} &=
          \B_n \,,
          \\
        \widetilde\B^{(x_j)}_0 &=
          \B^{(x_j)}_0
          +
          x_j^k
          \left[\C_0^{(x_j)}\widehat\G - \widehat\G\A_0^{(x_j)}\right] \,,
          \\
        \widetilde\B^{(x_j)}_{n>0} &=
          \B^{(x_j)}_n \,,
      \end{align*}
      and
      \begin{align*}
        \widetilde\D^{(x_j)}_n &=
          \D^{(x_j)}_n
          +
          \sum\limits_{i=0}^k 
          x_j^i\binom{k}{k-i}
          \E^{(x_j)}_{n+k-i}\widehat\G \,,
          \\
        \widetilde\D_{n<k} &=
          \D_n
          +
          \sum\limits_{x_j\in S} 
          \sum\limits_{m=0}^{k-n-1}
          \sum\limits_{i=0}^{m}
          (-1)^{k-n-m-1}
          x_j^{k-n-i-1}
          \binom{k-m-1}{n}
          \binom{k}{k+i-m} 
          \E^{(x_j)}_i\widehat\G \,,
          \\
        \widetilde\D_{n\geq k} &=
          \D_n
          +
          \E_{n-k}\widehat\G \,.
      \end{align*}
    \subsection{$\epsilon$-Factorization}
      The transformation required in the $\epsilon$-factorization is $x$-independent.
      Hence, every coefficient matrix in \eqref{eq:M} transforms the same and the transformation rule \eqref{eq:trans} becomes a similarity transformation
      \[
        \widetilde\M(x,\epsilon)
        =
        \T^{-1}(\epsilon)
        \M(x,\epsilon)
        \T(\epsilon)\,.
      \]
      The matrices $\T(\epsilon)$ and $\T^{-1}(\epsilon)$ have the form
      \[
        \T(\epsilon) = \begin{pmatrix}
          \mathds{1} & 0 & 0 \\
          0 & \widehat\T(\epsilon) & 0 \\
          0 & 0 & \mathds{1}
        \end{pmatrix}\,, \quad
        \T^{-1}(\epsilon) = \begin{pmatrix}
          \mathds{1} & 0 & 0 \\
          0 & \widehat\T^{-1}(\epsilon) & 0 \\
          0 & 0 & \mathds{1}
        \end{pmatrix}\,,
      \]
      with $\widehat\T(\epsilon)$ and $\widehat\T^{-1}(\epsilon)$ corresponding to block $\C$.
      Using the block-triangular structure \eqref{eq:triangular} yields
      \[
        \begin{pmatrix}
          \widetilde\A(x,\epsilon) & 0 & 0 \\
          \widetilde\B(x,\epsilon) & \widetilde\C(x,\epsilon) & 0 \\
          \widetilde\D(x,\epsilon) & \widetilde\E(x,\epsilon) & \widetilde\F(x,\epsilon)
        \end{pmatrix}
        =
        \begin{pmatrix}
          \A(x,\epsilon) & 0 & 0 \\
          \T^{-1}(\epsilon) \B(x,\epsilon) & \T^{-1}(\epsilon) \C(x,\epsilon) \T(\epsilon) & 0 \\
          \D(x,\epsilon) & \E(x,\epsilon) \T(\epsilon) & \F(x,\epsilon)
        \end{pmatrix}\,.
      \]
      Thus, besides block $\C$, block $\B$ and block $\E$ are influenced as well.

  \bibliography{epsilon}
\end{document}